\documentclass[twocolumn,preprintnumbers,amsmath,amssymb,superscriptaddress,floatfix]{revtex4}

\usepackage{amsmath}
\usepackage{amssymb}
\usepackage{mathtools}
\usepackage{booktabs}
\usepackage{tikz}

\def\quad{\hskip1em\relax}
\def\qquad{\hskip2em\relax}

\usepackage{enumerate}
\newcommand{\eps}{\epsilon}
\newcommand{\be}{\begin{equation}}
\newcommand{\ee}{\end{equation}}
\newcommand{\beno}{\begin{equation*}}
\newcommand{\eeno}{\end{equation*}}
\newcommand{\ba}{\begin{eqnarray}}
\newcommand{\ea}{\end{eqnarray}}
\newcommand{\bano}{\begin{eqnarray*}}
\newcommand{\eano}{\end{eqnarray*}}
\newcommand{\rar}{\rightarrow}

\usepackage{graphicx}

\newcommand{\YES}{\mathrm{YES}}

\begin{document}

\title{Notes on stochastic (bio)-logic gates: computing with allosteric cooperativity}
\author{Elena Agliari}
\affiliation{Dipartimento di Fisica, Sapienza Universit\`a di Roma, Italy}
\author{Matteo Altavilla}
\affiliation{Dipartimento di Matematica, Sapienza Universit\`a di Roma, Italy}
\author{Adriano Barra}
\affiliation{Dipartimento di Fisica, Sapienza Universit\`a di Roma, Italy}
\author{Lorenzo Dello Schiavo}
\affiliation{Dipartimento di Matematica, Sapienza Universit\`a di Roma, Italy}
\author{Evgeny Katz}
\affiliation{Department of Chemistry and Biomolecular Science, Clarkson University, New York, USA}

\begin{abstract}
\vskip .7truecm
Recent experimental breakthroughs have finally allowed to implement in-vitro reaction kinetics (the so called {\em enzyme based logic}) which code for two-inputs logic gates and mimic the stochastic AND (and NAND) as well as the stochastic OR (and NOR). This accomplishment, together with the already-known single-input gates (performing as YES and NOT), provides a logic base and paves the way to the development of powerful biotechnological devices. The investigation of this field would enormously benefit from a self-consistent, predictive, theoretical framework. Here we formulate a complete statistical mechanical description of the Monod-Wyman-Changeaux allosteric model for both single and double ligand systems, with the purpose of exploring their practical capabilities to express logical operators and/or perform logical operations. Mixing statistical mechanics with logics, and quantitatively   our findings with the available biochemical data, we successfully revise the concept of cooperativity (and anti-cooperativity) for allosteric systems, with particular emphasis on  its computational capabilities, the related ranges and scaling of the involved parameters and its differences with classical cooperativity (and anti-cooperativity).

\end{abstract}

\maketitle

\vskip 1cm

\section{Introduction}
\label{intro}

Cell's life is based on a hierarchical and modular organization of interactions among its molecules \cite{barabasi2}: a functional module is defined as a discrete ensemble of reactions whose functions are {\em separable} from those of other molecules. Such a separation can be of spatial origin (processes ruled by short range interactions) or of chemical origin (processes requiring specific interactions) \cite{hopfield-new}. The latter, i.e., chemical specificity, is at the basis of biological information processing \cite{prl1,prl2}. A paradigmatic example of this is the signal transduction pathway of the so called {\em two signal model} in immunology by which an effector  lymphocyte needs two signals (both integrated on its membrane's highly-specific receptors in a close temporal interval) to get active \cite{goodnow}: these signals are the presence of the antigen
\emph{and} the consensus of an helper-cell;
this constitutes a marvelous, biological, and stochastic, AND gate \cite{gino}.
We added the adjective {\em stochastic} because, quoting Germain, ``as one dissects the immune system at finer and finer levels of resolution, there is actually a decreasing predictability in the behavior of any particular unit of function", furthermore, ``no individual cell requires two signals (...) rather, the probability that many cells will divide more often is increased by co-stimulation" \cite{germain}.

Beyond countless natural examples, biologic gates have been realized even experimentally, see e.g. \cite{katz-or, katz-and, katz-general, Graham-PhysBiol2005,Prehoda-CurrOpinCellBiol2002,Tamsir-Nature2011,Kramer-BiotechBioeng2004,Setty-PNAS2003,Guet-Science2002,Dueber-CurrOpinStructBiol2004,Dueber-Science2003}, the ultimate goal being the experimental realization of stochastic, yet controllable, biological circuits \cite{ABBDU-ScRep2013,Infochemistry,Seeling-Science2006,Zhang-Science2007}.

Such striking outcomes also arouse a great theoretical attention aimed to develop a self-contained framework able to highlight their potentialities and suggest possible developments. In particular, statistical mechanics has proved to be a proper candidate tool for unveiling biological complexity: in the past two decades statistical mechanics has been applied to investigate intra-cellular (e.g. me\-ta\-bo\-lo\-mics \cite{meta,enzo}, proteinomics \cite{prote,prote2}) as well as extra-cellular (e.g. neural networks \cite{Coolen,neuro2}, immune networks \cite{immuno1,immuno2}) systems. Also, statistically mechanics intrinsically offers a partially-random scaffold which is the ideal setting for a stochastic logic gate theory.
\newline
Another route to unveil the spontaneous information processing capabilities of biological matters is naturally constituted by information theory and logics (see e.g. \cite{bio2,bio1} and references therein).

Remarkably, statistical mechanics and information theory (see the seminal works by Khinchin \cite{chi1,chi2}, and by Jaynes \cite{Jaynes1,Jaynes2}) and, in turn, information theory and logics (see the seminal works by Von Neumann \cite{neuman}, and by Chaitin \cite{chaitin}) have been highlighted to be deeply connected. Therefore, it is not surprising that even in the quantitative modeling of biological phenomena these two routes are not conflicting but, rather, complementary.
\newline
In this work, we will use the former (statistical mechanics) to describe a huge variety of biochemical allosteric reactions, and then, through the latter (mathematical logic), we will show how these reactions naturally encode stochastic versions of boolean gates and are thus capable of noisy information processing.

We will especially focus on allosteric reactions (as those of Koshland, Nemethy and Filmer (KNF) \cite{KNF} and  Monod-Wyman-Changeaux (MWC) \cite{MWC}) as they play a major role in enzymatic processes for which a great amount of experimental data is nowadays available.
However, classical reaction kinetics (i.e. those coded by Hill, Adair, etc. \cite{Hill}) can also perform logical calculations and along the paper we will deepen the crucial differences between the two types of kinetics -{\em allosteric cooperativity} versus {\em standard cooperativity}- when framed within a statistical mechanical scaffold.

Moreover, focusing primarily on the paradigmatic MWC model as a test case, we show that imposing the correct scalings and bounds on the involved parameters, gives rise to constraints which, if not properly accounted, may possibly prevent the system to perform as a logic gate.
%
%
%
%
%
%
\section{Results}
In the case of allosteric receptors, several models have been introduced. Many of these assume that a receptor can exist in either an active or inactive state, and that binding of a ligand changes the receptor bias to each state. In particular, in the Monod-Wyman-Changeaux (MWC) model, ligand-bound receptors can be in either state, but coupled receptors switch between states in synchrony. Beyond that pioneering work, several models able to provide qualitative and quantitative descriptions of binding phenomena have been further introduced in the Literature, as e.g. the sequential model by Koshland, Nemethy and Filmer (KNF).

Here we consider
MWC-like kinetics, and we try to map it into a statistical mechanical scaffold. We start by introducing terminology and parameters for mono-receptor/mono-ligand systems (playing for single input gates as YES and NOT) and then we expand such a scenario in order to account for the kinetics of more complex systems (double-receptors/double-ligands, as those will play for two-input gates as AND, NAND, OR, NOR).

The plan is as follows: Once introduced the microscopic settings (e.g., the occupancy states $\sigma_i$, $i \in (1,...,n)$ of $n$ receptors and the dissociation energy $h$), we define Hamiltonian functions $H_n(\sigma,h)$ coding for the chemical bindings; then -being $\beta$ the thermal noise $\beta=1/k_B T$ (where $k_B$ is the Boltzmann constant and $T$ represents the temperature) - we build their related Maxwell-Boltzmann probabilistic weights $\propto \exp[-\beta H_n(\sigma,h)]$; with the latter we can compute the partition functions $Z=\sum_{\sigma}\exp(-\beta H)$, both for the active state $Z_A$ and for the inactive $Z_I$ state.
\newline
Their ratios, $p_A=Z_A/(Z_A+Z_I)$ and $p_I=Z_I/(Z_A+Z_I)$ then return the probabilities of the active/inactive states as functions of the parameters (e.g. $\beta,h,n$).
\newline
These probabilities are first analyzed from a logic perspective in order to show how they can account for boolean gates and, then, used to successfully fit the outcomes of the experiments of enzyme based logic.
This route, although rather lengthy,  shows why allosteric mechanisms share similar behaviors with those of classical cooperativity, but, at the same time, clearly reveals deep differences between these phenomena.
\subsection{System description.} \label{sec:Model}
\begin{figure}[htb!]
\begin{center}
\includegraphics[scale=0.3]{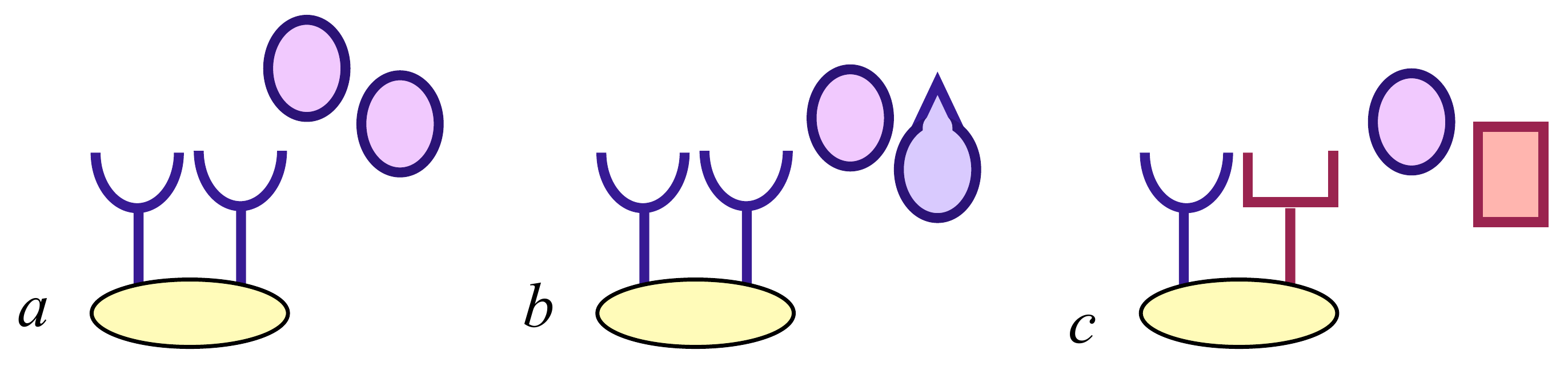}
\caption{This scheme summarizes the kind of systems we are considering here: Mono-receptor/Mono-ligand ($a$), Mono-receptor/Double-ligand ($b$) and Double-receptor/Double-ligand ($c$). In this cartoon all molecules are shown as dimeric, but cases $a$ and $b$ also work with monomeric structures. In the Mono-receptor/Mono-ligand case only one kind of receptor and one kind of ligand (compatible with the receptor) are considered; in the Mono-receptor/Double-ligand case we still have one kind of receptor, but two different ligands both compatible with the receptor; in the Double-receptor/Double-ligand case we consider molecules displaying two different receptors in the presence of two different ligands, each compatible with only one receptor.
\newline
The kinetics of these systems is addressed in Secs.~\ref{ssec:MM}, \ref{ssec:MB} and \ref{ssec:BB}, respectively while in Sec.~\ref{sec:Logic} they are shown to work as YES, OR, and AND logic gates. See also \cite{Ronde-Biophys2012}.
}
\label{fig:schema0}
\end{center}
\end{figure}

Specifically, we start considering a system built of several molecules, each displaying one or more receptors.
Each receptor exhibits multiple binding sites where a ligand can reversibly bind, and which can exist in different states (i.e. active and inactive).
In general the receptors exhibited by a given molecule can differ in e.g., the number of binding sites, the affinity with ligands, etc..
%
%
As we are building a theory for single and double input gates, in the following, we will focus on simple systems where receptors can house only one or two kinds of binding sites, as exemplified in Fig.~\ref{fig:schema0}.

The simplest system we consider is made of a set of receptors of the same kind and in the presence of a unique ligand (see panel $a$ in Fig.~\ref{fig:schema0}).
%
More precisely, each receptor is constituted by $n$ functionally identical bin\-ding sites
indexed by $i$, whose occupancy is given by a boolean vector $\sigma = \{ \sigma_i \}$, $i=1,...,n$ where $\sigma_i=1$ (respectively $0$) indicates the binding site $i$ is {\em occupied} (respectively {\em vacant}).

As required by the \emph{all-or-none} MWC model, a receptor is either {\em active} (T) \emph{or} {\em inactive} (R); the receptor state is indicated by a boolean {\em activation parameter} $a$, \ ($a =0,1$) \cite{Thom,Ronde-Biophys2012}.

In the \emph{absence of the ligand}, the active and inactive states (which are assumed to be in equilibrium) differ in their chemical potential, whose delta, indicated by $E$, can, in principle, be either positive (favoring the inactive state) or negative (favoring the active state).

Given a system of receptor molecules in the \emph{absence of ligand} and in equilibrium at a given temperature $T$, we pose the following assumptions:
\begin{enumerate}[$(a)$]
\item\label{assa} As both the active and inactive state may coexist, the composition of the system also depends on the parameter $L \equiv L(T)>0$, namely the \emph{equilibrium constant} at temperature $T$.  Letting $[R]$ be the total concentration of the receptors, $[R_A]$ (respectively $[R_I]$) the concentration of the \emph{active} (respectively \emph{inactive}) receptors \emph{in absence of the ligand}, it is $[R]=[R_A]+[R_I]$ and $[R_A]=L[R_I];$
\item\label{assd} For the sake of  simplicity, binding sites of a mono-receptor are considered as functionally identical (as in the original model \cite{MWC}).
\end{enumerate}

In the absence of ligand, we also need to establish which of the two states (namely the \emph{active} and \emph{inactive} one) has a higher chemical potential. As shown in the Literature (see \cite{Thom} and below) the choice is in general  arbitrary (i.e. case dependent), hence we take both possibilities into account. We therefore consider two sets of \emph{mutually exclusive} assumptions (the latter of which is denoted by a ``prime'' symbol).
\begin{enumerate}[$(a)$]
\setcounter{enumi}{2}
\item\label{assc} The active state has a \emph{higher} chemical potential \footnote{Notice that, while this assumption is in contrast with the original MWC model \cite{MWC}, the model itself is still self consistent as thoroughly explained in \cite{Thom}. The same conclusion may be drawn by the fact that, in the MWC paper, the opposite assumption is merely exploited for calculations.
} (i.e. $E>0$), as e.g. in \cite{Mello-PNAS2005}, \cite{Thom},  hence  the inactive state must then be predominant (to minimize energy) (i.e. $L\ll 1$);
\[\mathrm{AUT}\]
\item[$(c')$]\label{asscP}
The active state has a \emph{lower} chemical potential (i.e. $E'\equiv -E<0$) as e.g. in the original MWC model \cite{MWC}, hence (still for minimum energy requirement) the active state must then be predominant (i.e. \mbox{$L\gg 1$}).
\end{enumerate}

For a thorough comparison of these two alternative assumptions (and those of the original MWC) we refer to Tab.~\ref{tab1}.

For the sake of clarity we will from now on refer to the \eqref{assc}-type assumptions as ``assumptions $\mathcal{A}$'' and to the $(c')$-type assumptions as ``assumptions $\mathcal{A}^{\prime}$''. We also refer to the $\mathcal{A}^{\prime}$-set of assumptions as \emph{dual} to assumptions $\mathcal{A}$, where this terminology is introduced to match the one of mathematical logic and will be therefore explained in Sec.\ref{sec:Logic}.
All assumptions without a dual one are taken to be part of both the assumptions' sets.

Let us now discuss the case of a system of receptor molecules in the \emph{presence of ligand}. Clearly, the behavior of the system is expected to depend on ligand's  concentration $[S]$ \emph{and} on the receptor state (i.e. either active or inactive). The dependence on the receptor state is formalized by introducing \emph{dissociation constants} $K_A$ and $K_I$ for the receptor in the \emph{active} and \emph{inactive} state, respectively (see \cite{Ronde-Biophys2012}). Letting $[(R_AS)_i]$ be the concentration of the receptor/ligand complex's molecules which have exactly $i$ occupied binding sites, we define the average concentration of the \emph{active} receptor/ligand complex as
%
\beno
\langle [R_A \, S]\rangle\equiv \frac{1}{n}\sum_{i=1}^ni\cdot[(R_A \, S)_i].
\eeno
We can define the average concentration $\langle[R_I \, S]\rangle$ of the \emph{inactive} receptor/li\-gand complex in an analogous way, and we can then set
\beno
K_A\equiv \frac{[S][R]}{\langle[R_AS] \rangle},\ \ \ \  K_I \equiv \frac{[S][R]}{\langle[R_IS]\rangle},
\eeno
in accordance with the original presentation of MWC model \footnote{In \cite[p. 90]{MWC}, \emph{microscopic dissociation constants of a ligand} [...] \emph{bound to a stereospecific site} are considered, whose arithmetic weighted means we denote as global dissociation constants.}.
The dynamics of the receptor/ligand system is therefore determined by the variable $[S]$ and the parameters $K_A, K_I$.
\newline
Now, considering both the ligand and the receptor/ligand solution we assume that
\begin{enumerate}[$(a)$]
\setcounter{enumi}{3}
\item\label{asse} receptor-ligand solution is homogeneous and isotropic. This mean-field-like assumption is actually a key assumption of all the approaches in modeling classical reaction kinetics, see e.g. \cite{ABBDU-ScRep2013}.
\end{enumerate}

Finally, we address another (apparently) arbitrary choice, to answer the following question: is the ligand an \emph{activator}, i.e. its the presence enhances receptor's activation, or rather a \emph{suppressor}, i.e. its presence hindrances activation?

As it will be clear from Sec.~\ref{ssec:MM}, this choice is dual to and fully determined by the one made for chemical potential (assumptions \eqref{assc}'s). Indeed, to avoid \emph{trivial} (i.e. \emph{static}) behavior of the system, we have to set either
\begin{enumerate}[$(a)$]
\setcounter{enumi}{4}
\item\label{assf} The ligand is an \emph{activator}, i.e. the presence of the ligand enhances activation of the receptor. Therefore, the occupation of each receptor singularly \emph{decreases} the energy required for activation by a parameter \mbox{$\eps>0$}. 
\[\mathrm{AUT}\]
\item[$(e')$]\label{assfP} The ligand is a \emph{suppressor}, i.e. the presence of the ligand hindrances activation of the receptor. Therefore, the occupation of each receptor singularly \emph{increases} the energy required for activation by a parameter $\eps>0$.
\end{enumerate}

\begin{figure}[htb!]
\includegraphics[scale=0.25]{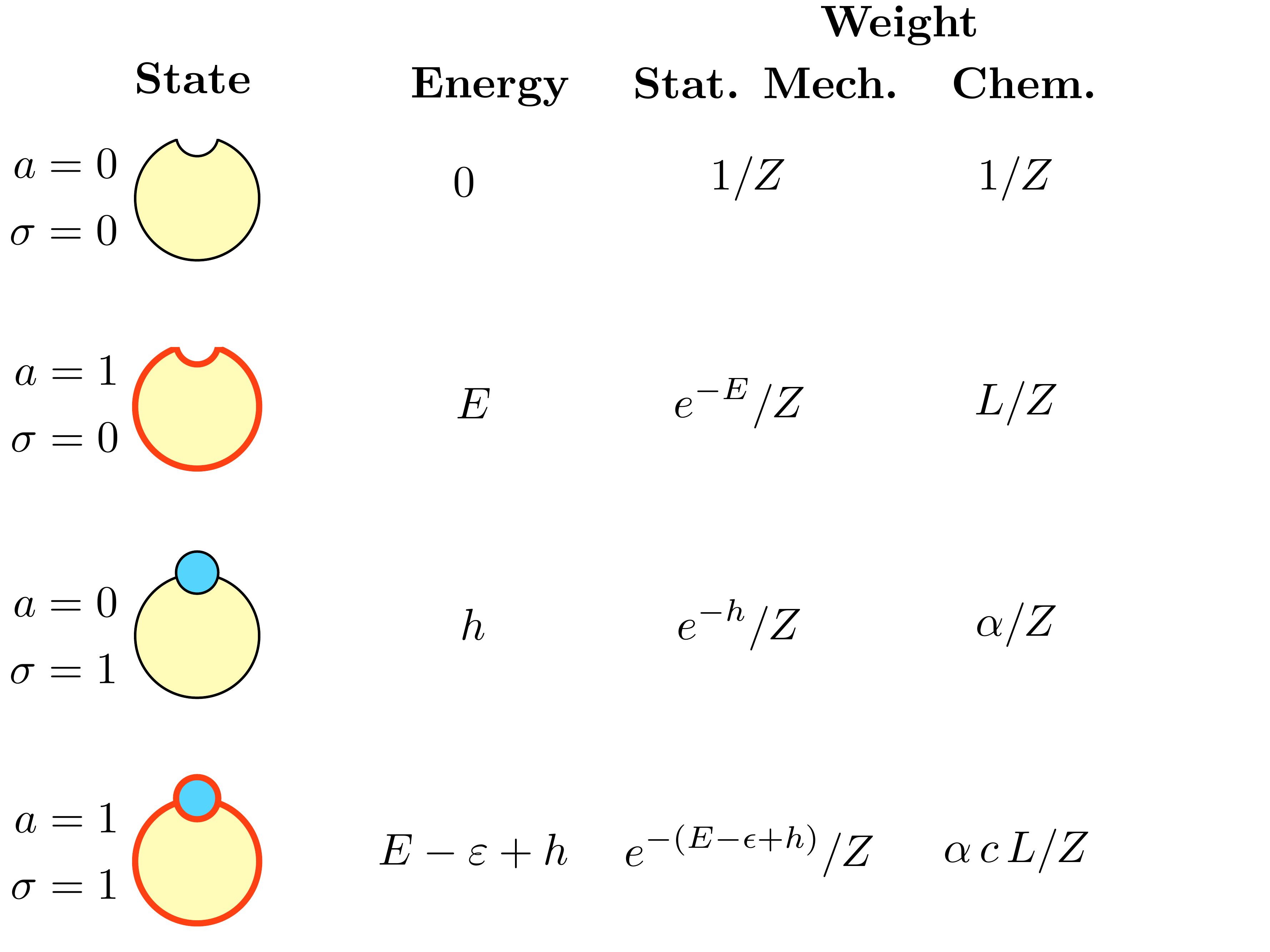}
\caption{This scheme summarizes the states and the weights of the simplest MWC molecule (that, in turn, codes for the YES gate). Having only one binding site ($n=1$), the number of possible states is four, from top to bottom: inactive and vacant, active and vacant, inactive and occupied, active and occupied. Each state corresponds to an energy, to a statistical mechanics weight and to a chemical weight. The energy is obtained by considering both the conformational degree of freedom of the molecule and the free energy of the binding process. The statistical mechanics weight is obtained according to the Boltzmann factor and the chemical weight is derived according to Tab.~\ref{tab1}. See also \cite{Marzen-JMB2013}.}
\label{fig:schema}
\end{figure}
\begin{table}[htb]\footnotesize
\caption{\footnotesize{Correspondences with the parameters originally used in \cite{MWC}}}
\label{tab1}
\begin{tabular}{ccc p{5cm}}
\toprule
\textbf{\textsf{Stat-Mech $\mathcal{A}$-set}} & \cite{Ronde-Biophys2012} & \cite{MWC} & \textbf{\textsf{MWC meaning}}\\
\midrule
$e^{-E}$ & $\omega_0$ & $L$& equilibrium constant of ac\-tive/in\-ac\-tive--state receptor system in absence of the ligand\\
$e^\eps$ & $\frac{K_I}{K_A}$ & $c$& dissociation constants ratio \\
$e^{-h}$ & $\frac{[S]}{K_I}$ & $\alpha$& neat percentage activation enhancement\\
$p_I$ &$\overline{R}$& & probability of the inactive (\emph{relaxed}) state, i.e. average concentration of the receptor in the inactive  state\\
$p_A$&$p_A$&$\overline{T}$ & probability of the active (\emph{tense}) state, i.e. average concentration of the receptor in the active state\\
\bottomrule
\end{tabular}
\begin{tabular}{ccc p{5cm}}
\toprule
\textbf{\textsf{Stat-Mech, $\mathcal{A}^{\prime}$-set}} & \cite{Ronde-Biophys2012} & \cite{MWC} & \textbf{\textsf{MWC meaning}}\\
\midrule
$e^{-E'}$ & $\omega_0$ & $L$& equilibrium constant of ac\-tive/in\-ac\-tive\--state receptor system in absence of the ligand\\
$e^{-\eps}$ & $\frac{K_A}{K_I}$ & $c^{-1}$& (inverse) dissociation constants ratio \\
$e^{h}$ & $\frac{K_I}{[S]}$ & $\alpha^{-1}$& inverse neat percentage activation enhancement\\
$p_I$ &$\overline{R}$& & probability of the inactive (\emph{relaxed}) state, i.e. average concentration of the receptor in the inactive  state\\
$p_A$&$p_A$&$\overline{T}$ & probability of the active (\emph{tense}) state, i.e. average concentration of the receptor in the active state\\
\bottomrule
\end{tabular}
\end{table}
%
%
%
%
%
\subsubsection{Mono receptor/Mono ligand (MM) properties at equilibrium.}\label{ssec:MM}
Under assumptions $\mathcal{A}$, any mono-receptor/mono-ligand system, built by $n$ receptors $[i \in (1,...,n)]$, and whose occupancy is ruled by $\sigma_i = (0,1)$, can be described by the following allosteric Hamiltonian function
\be\label{yeshamiltonian}
H(\sigma,a) =  \left( E-\eps\sum_{i=1}^n\sigma_i \right) a+h\sum_{i=1}^n\sigma_i,
\ee
where we recall $E$ to be the energy delta given by che\-mi\-cal potential, and we define $h$ to be the dissociation e\-ner\-gy, namely the energy captured by a single binding site of the inactive state receptor by binding to a ligand molecule \footnote{By definition, the dissociation energy introduced within this `physical framework' is related to the ligand concentration $[S]$ and $K_I$, as the latter enhances, or hindrances, the capability of a ligand's molecule to bind.};
the term in the brackets accounts for the fact that ligand acts as an activator since, for the active state ($a=1$) binding is energetically favored, while in the inactive state ($a=0$) the related term disappears in the Hamiltonian that reduces to the last term accounting for the association energy.

By the same reasoning under assumptions $\mathcal{A}^{\prime}$, we obtain
\be\label{nothamiltonian}
H(\sigma,a) =  \left(-E+\eps\sum_{i=1}^n\sigma_i \right) a+h\sum_{i=1}^n\sigma_i,
\ee

The main features of the mono-receptor/mono-ligand system described above are summarized in Fig.~\ref{fig:schema}.

It is worth highlighting that the Hamiltonians \eqref{yeshamiltonian} and \eqref{nothamiltonian} do not include any two-body couplings, i.e. any term $\propto \sum_{ij} \sigma_i \sigma_j$: this framework is intrinsically \emph{one-body} in the statistical mechanical vocabulary and this has implications in biochemistry too. For instance one-body theories do not undergo phase transitions, and, as the latter mirror ultra-sensitive reactions in chemical kinetics \cite{ABBDU-ScRep2013}, those are ruled out by this formalism.


Since the activation parameter is boolean, the receptor/ligand complex state may be considered regardless of the state of the receptor, by introducing the two Hamiltonians $H_A(\sigma) \equiv H(\sigma,1)$ and $H_I(\sigma)\equiv H(\sigma,0)$, defining the \emph{active} and the \emph{inactive} state energy, respectively. The corresponding partition functions are
\bano \label{yespartitions}
Z_A &=& \sum_{\{\sigma\}}e^{-\beta H_A(\sigma)} \\
Z_I &=& \sum_{\{\sigma\}}e^{-\beta H_I(\sigma)},
\eano
while the total partition function $Z$ 
is given by
\be \label{eq:3}
Z=\sum_{\{\sigma\},\{a\}} e^{-\beta H(\sigma,a)}=Z_A+Z_I.
\ee
A few remarks are in order here:
\newline
$-$ The summations in the partition function (\ref{eq:3}) account for the activation degree of freedom too. This means that the latter participate in thermalization or, in other words, that the intrinsic timescale for the dynamics of $a$ is bounded from above by those of the $\sigma$: this is consistent with the original MWC assumptions of synchronized switches among coupled receptors (the so called {\em all-or-none} behavior).
\newline
$-$ This model can be solved even {\em at finite $n$}, namely without the oversimplifying thermodynamics limit $n\to\infty$\\
$-$ All the energies can be expressed in units of the thermal energy $k_B T \equiv \beta^{-1}$, hence, in order to avoid possible misunderstanding as $T$ already addresses the tense molecular state and to keep notation as simple as possible, in the following we set $\beta=1$, thus forcing all aforementioned parameters and variables to be dimensionless\\
$-$ As a consequence of the previous two remarks, the stochasticity is retained by the parameter $n$, such that for $n\to \infty$ stochastic computing will collapse on deterministic one (that of classical logic), while the smaller $n$, the larger the noise affecting the system.

Now, focusing on $Z_A$ (as $Z_I$ is analogous), we define $k = \sum_{i=0}^n \sigma_i$, and we can therefore write
\bano
 Z_A &=& \sum_{(\sigma,1)} e^{-H(\sigma,1)} = \\
 &=& \sum_{k=0}^n A_k e^{-(E-k\eps)-hk} = e^{-E} \sum_{k=0}^n A_k e^{k(\eps-h)},
\eano
where $A_k$ denotes the number of times that the sum $\sum_{i=1}^n\sigma_i$ turns out to be equal to $k$.
Noting that $\sigma$ is a binary vector, we get straightforwardly that $A_k = \binom{n}{k}$, and therefore
\bano
 Z_A &=& e^{-E} \sum_{k=0}^n \binom{n}{k} e^{k(\eps-h)} = e^{-E} \sum_{k=0}^n \binom{n}{k} e^{k(\eps-h)}\cdot 1^{n-k}\\
 &=&   e^{-E} [1+ e^{(\eps-h)}]^n.
\eano
Analogously, $Z_I = (1+ e^{-h})^n$.

Therefore, the probability $p_A$ and $p_I$ for the complex to be in the active and in the inactive state respectively are
\be\label{pAyes}
(p_A)_{MM} = \tfrac{Z_A}{Z_A+Z_I}=\tfrac{e^{-E}(1+e^{\eps-h})^n}{e^{-E}(1+e^{\eps-h})^n+(1+e^{-h})^n},
\ee
\be
\nonumber
(p_I)_{MM}  = \tfrac{Z_I}{Z_A+Z_I}=\tfrac{(1+e^{-h})^n}{e^{-E}(1+e^{\eps-h})^n+(1+e^{-h})^n}.
\ee
where the subscript $MM$ stands for ``Mono-Mono''.

Correspondingly,
\be\label{pAnot}
(p_A')_{MM} = \tfrac{Z_A}{Z_A+Z_I}=\tfrac{e^{E}(1+e^{-\eps-h})^n}{e^{E}(1+e^{-\eps-h})^n+(1+e^{-h})^n},
\ee
\be \nonumber
(p_I')_{MM}  = \tfrac{Z_I}{Z_A+Z_I}=\tfrac{(1+e^{-h})^n}{e^{E}(1+e^{-\eps-h})^n+(1+e^{-h})^n}.
\ee

The interesting quantity to look at is $(p_A)_{MM}$, as it corresponds to the concentration $\overline T$ of receptors in the active state and this is expected to continuously increase (respectively decrease) with the percentage of activation enhancement (i.e. $e^{-h}$, see Tab.~\ref{tab1}) under assumptions $\mathcal{A}$ (respectively $\mathcal{A}^{\prime}$).
We notice though that the original model \cite{MWC} is concerned with $\overline R$ (i.e. with $p_I$) rather than $\overline{T}$; anyhow, $p_A$ and $p_I$ carry the same information as they are complementary probabilities.
\newline
Notably, the correspondence stated in Tab.~\ref{tab1} confirms the consequences of assumptions (\ref{assc}) and (\ref{assf}), that is, choosing $E>0$ yields $L< 1$, while choosing $\eps>0$ yields $c> 1$.
In particular, according to the notation of \cite{MWC}, we have
\bano
\overline{R}&=&\frac{(1+\alpha)^n}{L(1+c\alpha)^n+(1+\alpha)^n},\\
\overline{T} &=& 1-\overline{R}=\frac{L(1+c\alpha)^n}{L(1+c\alpha)^n+(1+\alpha)^n}.
\eano
Conclusions on the dual assumptions $\mathcal{A}^{\prime}$ are much the same and will not be repeated.

%
%
%
%

\subsubsection{Mono-receptor/Double-ligand (MD) properties at equilibrium}  \label{ssec:MB}
Under the assumptions of the previous section, any mono-receptor/double-ligand system, built by $n$ receptors $[i \in (1,...,n)]$ and whose occupancy is ruled by $\sigma_i =(0,1)$, can be described by the following allosteric Hamiltonian function
\be\label{orhamiltonian}
H(\sigma, a,I,J) = \left( E- \eps \sum_{i=1}^n \sigma_i \right)a + h_1 \sum_{i \in I} \sigma_i + h_2 \sum_{j \in J} \sigma_j,
\ee
where, in contrast with the previous case described by eq.(~\ref{yeshamiltonian}), two distinct li\-gands, whose dissociation energies are denoted by $h_1$ and $h_2$ respectively, are considered. More precisely, $I$ and $J$ are two subsets of $\{1, \dots, n\}$ such that $I \cap J= \varnothing$, and they denote the sites linked to the first ligand and to the second ligand, respectively. As a condition to simulate this, we impose that
$I \cup J = \{ \text{indices such that } \sigma_i =1\}$.

As we did for the Mono-Mono case, the partition function coupled to the Hamiltonian (\ref{orhamiltonian}) is given by
\bano
  Z&=& \sum_{(\sigma,a)} e^{-H(\sigma,a)} = \sum_{(\sigma,0)} e^{-H(\sigma,0)} + \sum_{(\sigma,1)} e^{-H(\sigma,1)} \\
  &=& Z_I + Z_A.
\eano
We focus on $Z_A$, as $Z_I$ is analogous. Let us pose $k_1= |I|$ and $k_2= |J|$, notice that $k=k_1 +k_2= \sum_{i=0}^n \sigma_i$, and write the sums explicitly as
\beno
 Z_A = \sum_{(\sigma,1)} e^{-H(\sigma,1)} = \sum_{k=0}^n \sum_{k_1=0}^k A_{k,k_1} e^{-(E-k\eps)-h_1k_1-h_2(k-k_1)},
\eeno
where $A_{k,k_1}$ denotes the number of times that the sum $\sum_{i=0}^n \sigma_i$  is equal to $k$, with the condition that $k_1$ of the $\sigma_i$'s
belong to the set $I$. This quantity is rather tricky to calculate but can actually be rewritten in terms of multinomial coefficient (which counts the number of ways we can choose $k$ elements among $n$, with the condition that they are divided in groups of $k_j$ elements each).
Then, we get
\beno
 A_{k,k_1} = {n \choose k_1, k-k_1} = {n \choose k_1, k_2}
\eeno
in such a way that $Z_A$ can be rewritten (using $k_1$ and $k_2$) as
%
\begin{align*}\nonumber
 Z_A =& e^{-E} \sum_{k_1+k_2=0}^n \sum_{k_1=0}^{k_1+k_2} {n \choose k_1, k_2} e^{(k_1+k_2)\eps-h_1k_1-h_2k_2} \\
 \nonumber =& e^{-E}  \sum_{k_1+k_2=0}^n \sum_{k_1=0}^{k_1+k_2} {n \choose k_1, k_2} e^{k_1(\eps-h_1)}\cdot e^{k_2(\eps-h_2)} \\
\nonumber =& e^{-E} \left[1+ e^{(\eps-h_1)} + e^{(\eps-h_2)} \right]^n,
\end{align*}
where in the second passage we must consider a $1^{n-(k_1+k_2)}$ factor, which allows us to conclude the calculation,
by simply expanding the trinomial.
\newline
Analogously, we obtain $Z_I = (1+e^{-h_1}+e^{-h_2})^n$.

Indeed, we have
\be \label{pAor}
(p_A)_{MD} = \tfrac{e^{-E}(1+e^{\eps-h_1}+e^{\eps-h_2})^n}{e^{-E}(1+e^{\eps-h_1}+e^{\eps-h_2})^n+(1+e^{-h_1}+e^{-h_2})^n},
\ee
\be  \nonumber
(p_I)_{MD} = \tfrac{(1+e^{-h_1}+e^{-h_2})^n}{e^{-E}(1+e^{\eps-h_1}+e^{\eps-h_2})^n+(1+e^{-h_1}+e^{-h_2})^n}.
\ee

In a similar fashion, under assumptions $\mathcal{A}^{\prime}$ we obtain
\be \label{pAnor}
(p_A')_{MD} = \tfrac{e^{E}(1+e^{-\eps-h_1}+e^{-\eps-h_2})^n}{e^{E}(1+e^{-\eps-h_1}+e^{-\eps-h_2})^n+(1+e^{-h_1}+e^{-h_2})^n},
\ee
\be \nonumber
(p_I')_{MD}  = \tfrac{(1+e^{-h_1}+e^{-h_2})^n}{e^{E}(1+e^{-\eps-h_1}+e^{-\eps-h_2})^n+(1+e^{-h_1}+e^{-h_2})^n},
\ee
where the subscript $MD$ stands for ``Mono-Double''.


\subsubsection{Double-receptor/Double-ligand (DD) properties at equilibrium}  \label{ssec:BB}
Under the same assumptions of the previous sections, any double-receptor/double-ligand system, built by $n$ receptors $[i \in (1,...,n)]$ and whose occupancy is ruled by $\sigma_i =(0,1)$, can be described by the following allosteric Hamiltonian function
\ba \label{andhamiltonian}
&&H(\sigma, \tau ,a) =  H_1(\sigma,a) + H_2(\tau,a)\\
\nonumber
&&= \left( 2E-\eps_1 \sum_{i=1}^{n_1} \sigma_i -\eps_2 \sum_{i=1}^{n_2} \tau_i \right ) a + h_1 \sum_{i=1}^{n_1}\sigma_i + h_2 \sum_{i=1}^{n_2}\tau_i.
\ea
We note that the system factorizes into two independent Mono-Mono Hamiltonians, hence we can entirely skip the calculations, referring to results of Sec.~\ref{ssec:MM}. Thus, focusing on a symmetric case for simplicity, i.e. $\epsilon_1 = \epsilon_2 = \epsilon$ and $n_1=n_2=n$, we get for $(p_A)_{DD}$
\be \label{Pand}
\tfrac{e^{-2E} (1 + e^{\epsilon - h_1} + e^{\epsilon - h_2} + e^{\epsilon - h_1} e^{\epsilon - h_2} )^n}{ (1+e^{-h_1})^n (1 + e^{-h_2})^n +  e^{-2E}(1 + e^{\epsilon - h_1} + e^{\epsilon - h_2} + e^{\epsilon - h_1} e^{\epsilon - h_2} )^n},
\ee
while, via the dual assumptions $\mathcal{A}^{\prime}$, we have for $(p_A')_{DD}$
\be \label{PNAND}
\tfrac{e^{2E} (1 + e^{-\epsilon - h_1} + e^{-\epsilon - h_2} + e^{-\epsilon - h_1} e^{-\epsilon - h_2} )^n}{ (1+e^{-h_1})^n (1 + e^{-h_2})^n +  e^{2E}(1 + e^{-\epsilon - h_1} + e^{-\epsilon - h_2} + e^{-\epsilon - h_1} e^{-\epsilon - h_2} )^n}.
\ee
\subsection{Logical operations.} \label{sec:Logic}

Let us now explore the possibility of using these allosteric re\-cep\-tor-ligand systems as operators mimicking stochastic logic gates:  the \emph{presence} of ligands ({\em variables} in Logic) is denoted as $S_i$ for the $i$-th ligand, and the \emph{presence} of receptors ({\em operators} in Logic) is denoted as $R_{A,i}$ and $R_{I,i}$ for the active and inactive state of the $i$-th receptor, respectively \footnote{Note that $\sigma_i$ and $S_i$ are conceptually different because, in Logic, $S_i$ mirrors the presence of the $i$-{th} ligand, that is $S_i=$``true'' stands for a high concentration presence of the $i$-th ligand, thus within the statistical mechanical route the $S$'s are closer to the $h$'s than the $\sigma$'s.}.
\newline
Operators are of two kinds: the \emph{unary} operators YES and NOT, which evaluate a single argument, and the \emph{binary} operators, e.g., AND and OR, which evaluate two arguments.
\newline
Let us describe the examples of concrete interest in the paper:

$-$ \emph{Affirmation}: ``S'', namely the signaling of the \emph{presence} of ligand $S$. Hereafter this operator will be denoted as stochastic YES (or, in case a distinction between several ligands is necessary, as YES$_S$).

$-$ \emph{Negation}: ``$\neg S$'', namely the evaluation of the \emph{absence} of ligand $S$, which returns true if and only if the ligand $S$  is \emph{not} present. Hereafter this operator will be denoted as stochastic  NOT (or NOT$_S$).

$-$ \emph{Conjunction}: ``$S_1 \wedge S_2$'', namely the evaluation of the presence of \emph{both} ligands, which returns  ``true'' whenever both ligands occur to be present (i.e., in the case that $S_1$ and $S_2$ are assigned value ``true'') and ``false'' whenever at least one of the two ligands is not present (i.e., in the case that either $S_1$ or $S_2$ are assigned value false).
The evaluation of such operator is hereafter denoted as $S_1$ AND $S_2$ (stochastic  AND).

$-$  \emph{Non-exclusive disjunction}: $S_1 \vee S_2$, namely the evaluation of the presence of at least one ligand, which returns true whenever at least one ligand is present and value false whenever they are both absent.
The evaluation of such operator is hereafter denoted as $S_1$ OR $S_2$ (stochastic  OR).

As we will see, the receptor molecule plays as an operator, while ligands play as variables. In order to evaluate the formula, each variable can assume value either ``true''  of ``false'' according to the ligand concentration, where ``true" means that the ligand is present at a concentration larger than a threshold value, while ``false'' means that the ligand concetration is smaller than such a value. Moreover, the value arising from the evaluation of the operators corresponds to the activation state of the receptor: active if the evaluation returns ``true'' and inactive is evaluation returns ``false''.

\subsubsection{Mono-receptor/Mono-ligand system: YES and NOT functions.}
All the plots in this and the following sections are based on some scaling assumptions that will be discussed further in the paper (see Sec.~\ref{ssec:Scal}).
These assumptions are essential to our purpose (that is, they enable us to tune the free variables introduced defining the Hamiltonians), and are
deduced by physical and biochemical reasoning. We will refer to these assumptions as they are reported in Sec.\ref{sec:methods} below.

\begin{figure}[htb!]
\includegraphics[scale=0.35]{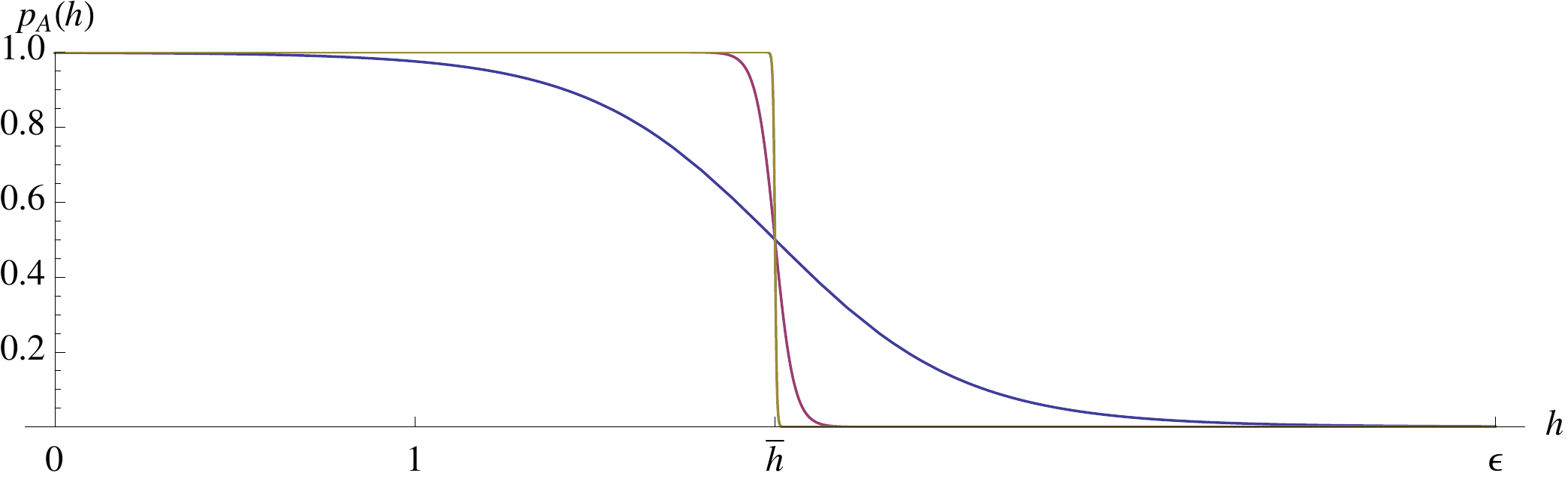}
\includegraphics[scale=0.35]{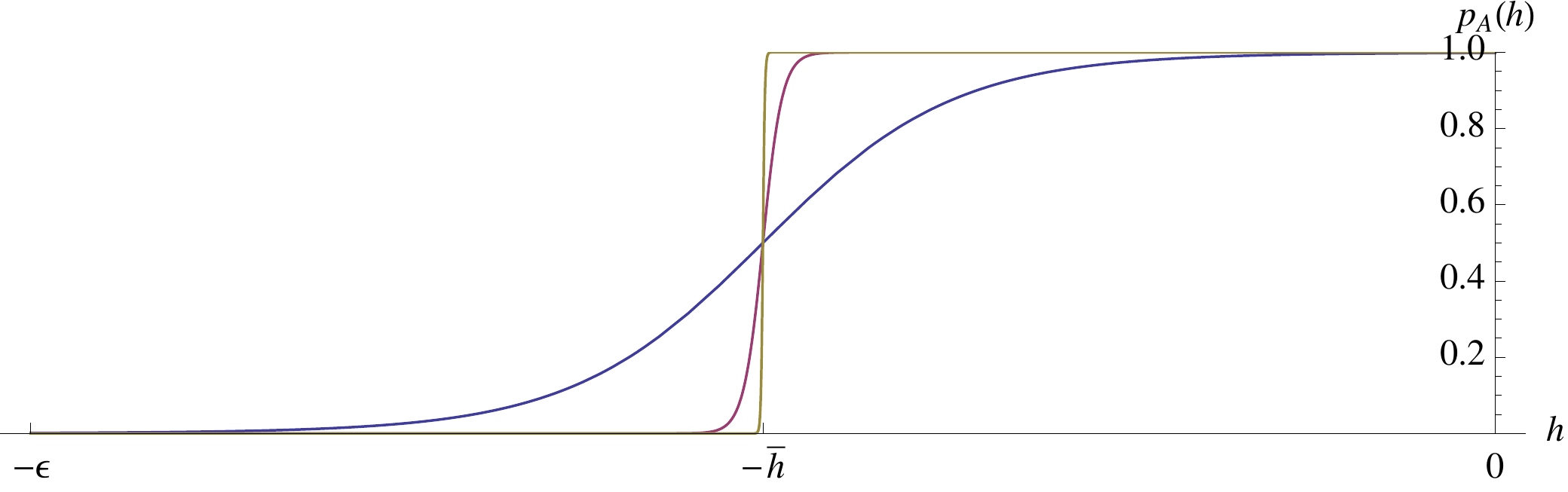}
\caption{Upper panel: Sigmoidal behavior of $p_A(h)$ with parameters $E=2n$, $\eps=2E/n$, where $n=5$ (\emph{blue}), $n=50$ (\emph{red}), $n=500$ (\emph{gold}). Lower panel: Anti-sigmoidal behavior of $p_A'(h)$ with parameters $E=2n$, $\eps=2E/n$, where $n=5$ (\emph{blue}), $n=50$ (\emph{red}), $n=500$ (\emph{gold}).}
\begin{center} \label{yesplot1D}
\end{center}
\end{figure}

Under scaling assumptions \eqref{hbioscaling}, \eqref{eq:qui} and \eqref{eq:Escal}, plots of the activation probability $(p_A)_{MM}(h)$
from eq.~\eqref{pAyes} show marked sigmoidal behavior (see Figure \ref{yesplot1D}, upper panel), signaling activation of the receptor in significative presence of the ligand, i.e. for small values of the variable $h$ \footnote{The logarithmic relation among $h$ and the concentration follows directly both from the original MWC model, as summarized in Table \ref{tab1} and the Thompson approach (see \cite{Thom}).}. 

Thus, the function $(p_A)_{MM}$ may be considered as mimicking the logical $\YES_{[L]}$ function, assuming boolean values $0$ for low ligand concentration and $1$ for high ligand concentration,
as one can see from Tab.~\ref{tab:POS}.

The threshold value is set at $\overline{h}$ which can in turn be fixed by properly choosing the system constituents (e.g. the number of binding sites hosted by a receptor).

On the contrary, the function $(p_A')_{MM}$ of eq.~\eqref{pAnot} may be considered as mimicking the logical NOT$_{[L]}$ function (Figure \ref{yesplot1D}, lower panel), assuming boolean values $0$ for high ligand concentration and $1$ for low ligand concentration, as one can see from Table \ref{tab:POS} below.

\subsubsection{Mono-receptor/Double-ligand system: OR and NOR functions}
The activation probability $(p_A)_{MD}$ (eq.~\eqref{pAor}) can be used to model a stochastic version of  the logic gate OR. In fact, if we look at the presence of the two different ligands as a binary input, the behavior of $(p_A)_{MD}$ (with the scaling assumptions of eqs.~\eqref{hbioscaling}, \eqref{eq:qui}), as a function of $h_1$ and $h_2$ (see Fig.~4), recovers the OR's one (see Tab. \ref{tab:POS}). Similarly to the YES case, the value 0 for $h_1$,\ $h_2$ denotes the saturation of the ligand. Therefore, consistently with the structure of OR, the presence of only one out of the two ligands is sufficient to make the molecule active; conversely, the value $\epsilon$ denotes the absence, thus for $h_1=h_2 \simeq \epsilon$, $(p_A)_{MD}$ is vanishing, namely, it returns as output ``false''.
\newline
Note that the projection of the plot over $h_1 =\eps$ (or $h_2=\eps$) gives a sigmoid, consistently with the fact that, if one of the two inputs is constantly false, the OR recovers the YES.

\begin{figure*}[htb!]\label{fig:OR}
\includegraphics[scale=0.35]{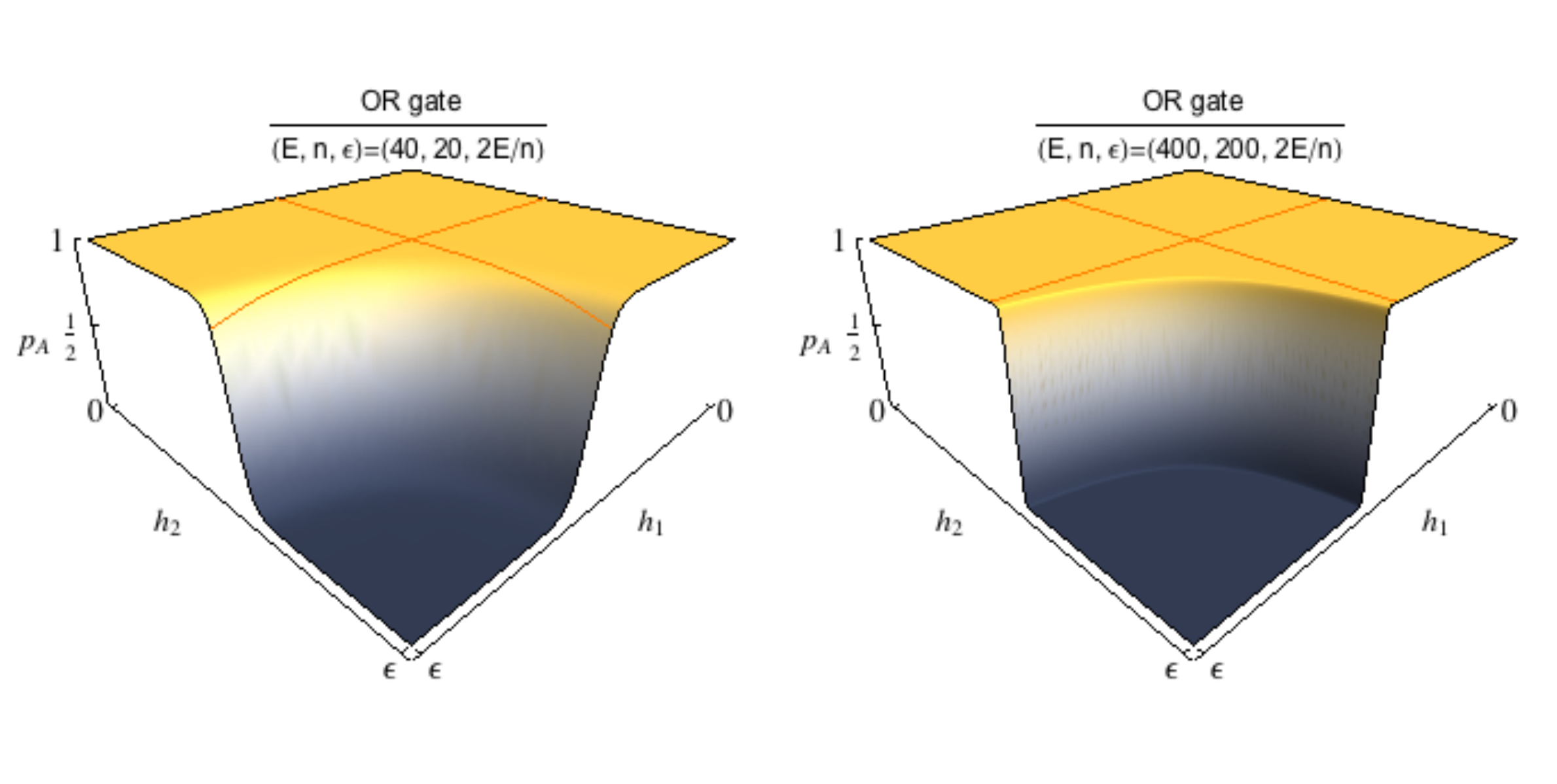}\includegraphics[scale=.35]{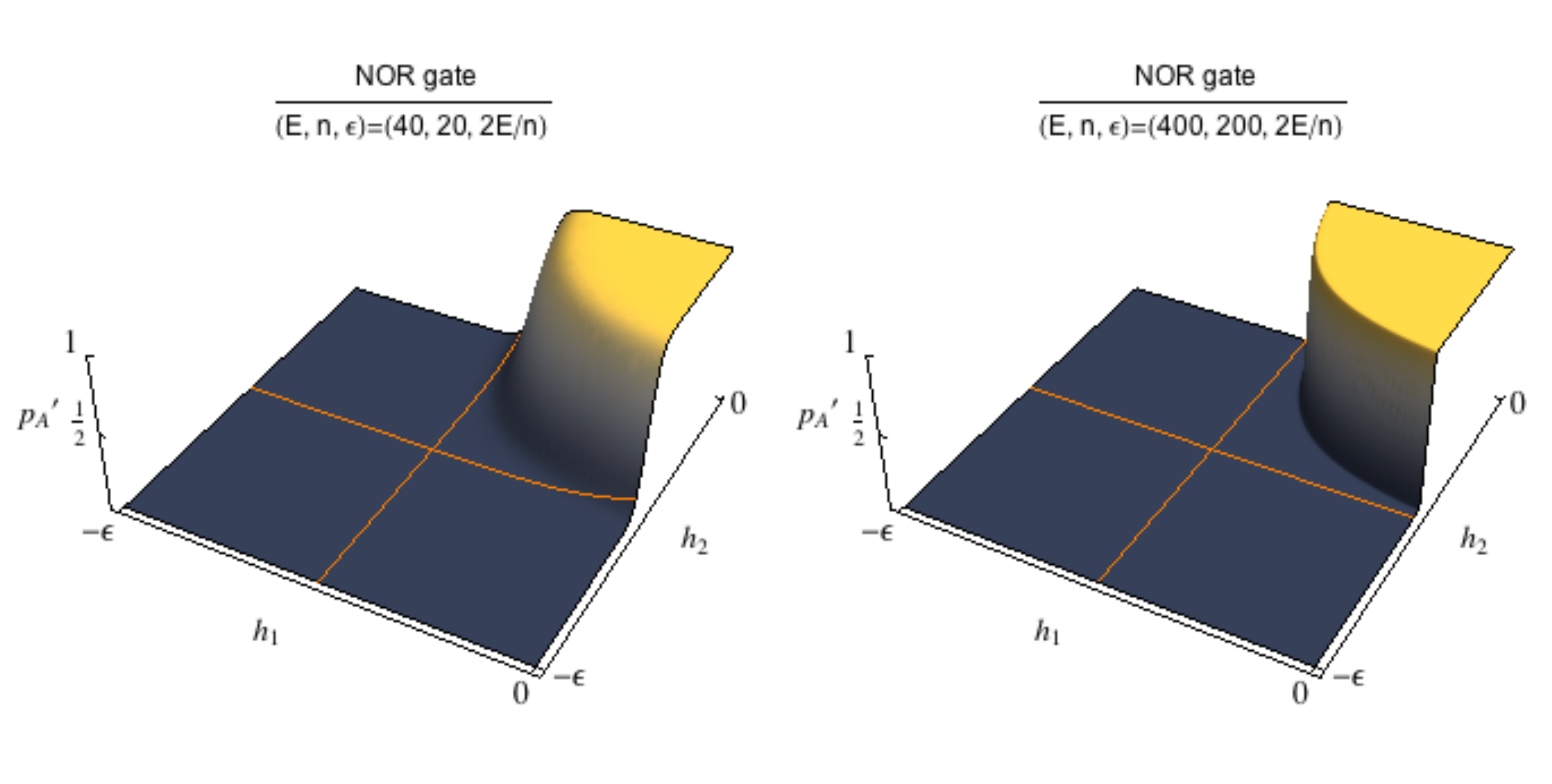}
\caption{Left: $(p_A)_{MD}(h_1,h_2)$ plots. Activation of the receptor is achieved by small values of $h_1$ or $h_2$, corresponding to a significative presence of any of the two ligands, thus simulating a stochastic OR function.
Right: $(p_A')_{MD}(h_1,h_2)$ plots. Activation of the receptor is verified by small values of $h_1$ or $h_2$, corresponding to a significative presence of any of the two ligands, thus simulating a stochastic NOR function. Note that for smaller $n$ curves are smooth (noisy), while for large $n$ quasi-discontinuous jumps appear.}
\end{figure*}

Performing the same calculations, the dual counterpart $(p_A')_{MD}$ of eq.~\eqref{pAnor} models the logical NOR gate, that is the direct negation of the previous one, as shown in Fig.~4.

\subsubsection{Double-receptor/Double-ligand system: AND and NAND functions.}

The function $(p_A)_{DD}$ described in Sec.~\ref{ssec:BB} (eq.~\eqref{Pand}) models a stochastic version of the logic AND gate (see Tab.~\ref{tab:POS}). As in the case of OR, we look at the two ligands as a binary input, and we assume the scaling assumptions coded in eqs.~\eqref{hbioscaling}, \eqref{eq:Escal2}, \eqref{eq:qui2}. The resulting behavior of  $(p_A)_{DD}$ fits 
the one expected for the AND function, with fitness to the expected plot that sensibly improves in the extremal regions of the plot, i.e. for $h_{1,2}\sim 0,\eps$ (see Fig.~5). Again, its projection  returns a sigmoid because if one of the two inputs is constantly true, the AND recovers the YES.

\begin{figure*}[htb!]\label{fig:AND}
\includegraphics[scale=.35]{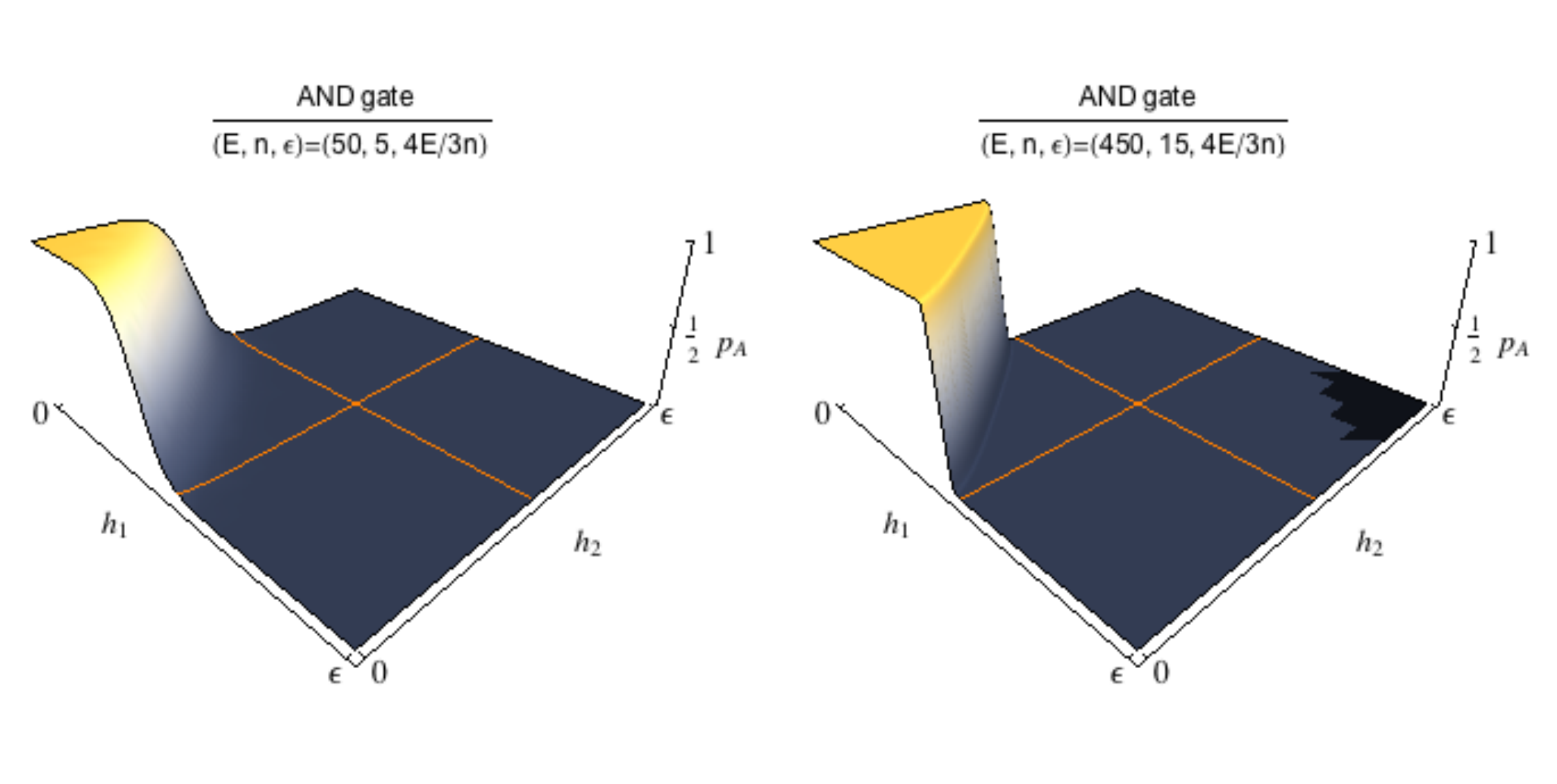}\includegraphics[scale=.35]{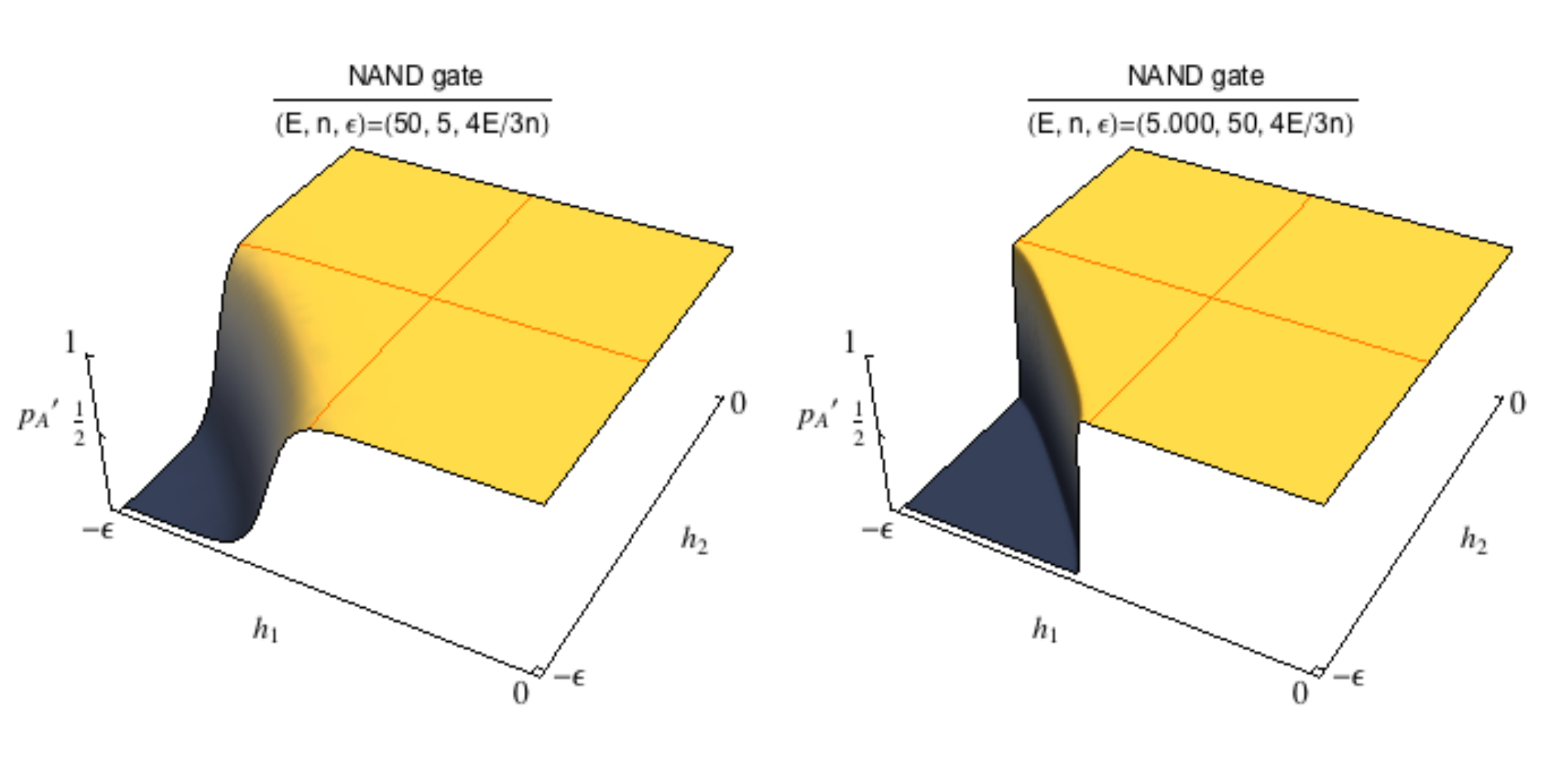}
\caption{Left: $(p_A)_{DD}(h_1,h_2)$ plots. Activation of the receptor is verified by small values of $h_1$ and $h_2$, corresponding to a significative presence of both the two ligands, thus simulating a stochastic AND function.
Right: $(p_A')_{DD}(h_1,h_2)$ plots. Activation of the receptor is verified by high (i.e. small in absolute value) values of $h_1$ or $h_2$, corresponding to a significative presence of any of the two ligands, thus simulating a stochastic NAND function.
\vspace{1cm}}
\end{figure*}

The dual version $(p_A')_{DD}$ (eq.~\eqref{PNAND}) models the logic gate NAND, i.e. the direct negation of the previous one. As this negation is precisely dual, so is the shape of the plot (see Fig.~5).


\begin{table}[htb]\footnotesize
\caption{\footnotesize{The truth table of all the logical operators introduced by now}}
\label{tab:POS}
\centering
\begin{tabular}{cc  c  c  c  c  c  c}
\toprule
\multicolumn{2}{c}{Input} & YES$_A$ & NOT$_A$ & $A$ OR $B$ & $A$ NOR $B$ & $A$ AND $B$ & $A$ NAND $B$ \\
A & B & $A$ & $\neg A$ & $A \vee B$ & $A\downarrow B$ & $A \wedge B$ & $A\uparrow B$\\
\midrule
1 & 1 & 1 & 0 & 1 & 0 & 1 & 0\\
1 & 0 & 1 & 0 & 1 & 0 & 0 & 1\\
0 & 1 & 0 & 1 & 1 & 0 & 0 & 1\\
0 & 0 & 0 & 1 & 0 & 1 & 0 & 1\\
\bottomrule
\end{tabular}
\end{table}


\begin{figure}[htb!]
\includegraphics[scale=0.6]{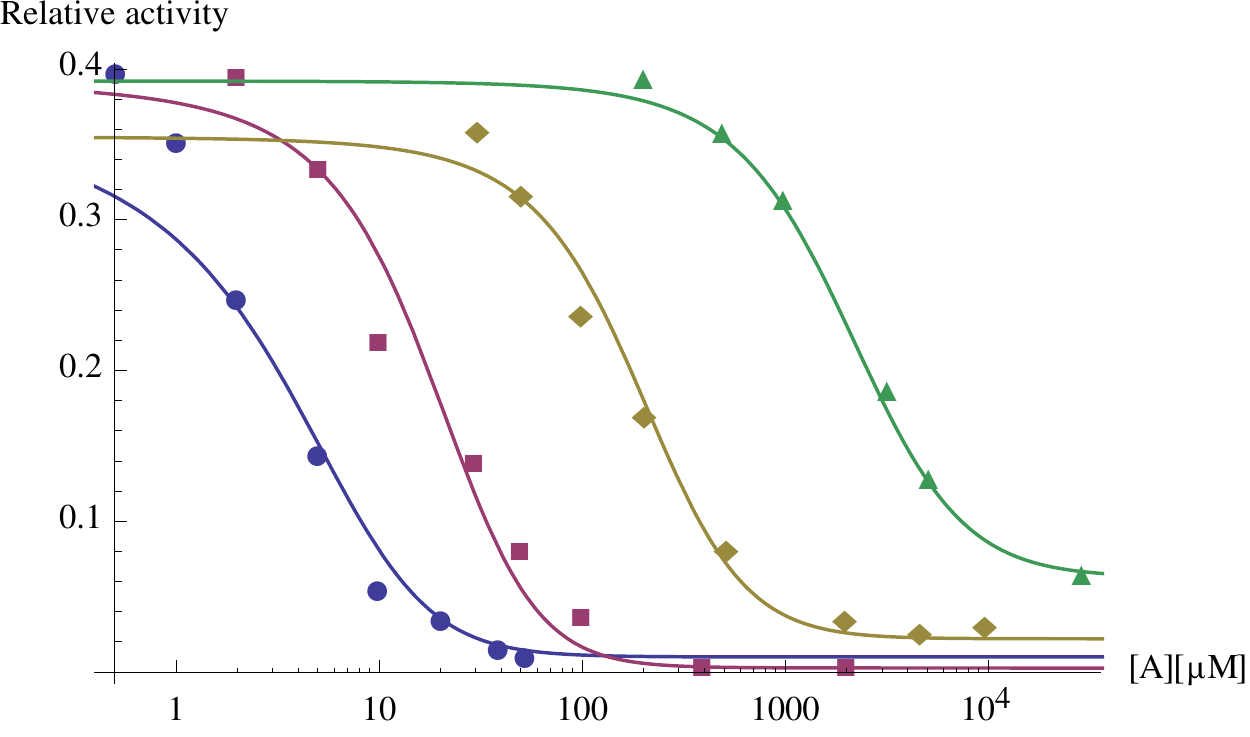}
\includegraphics[scale=0.6]{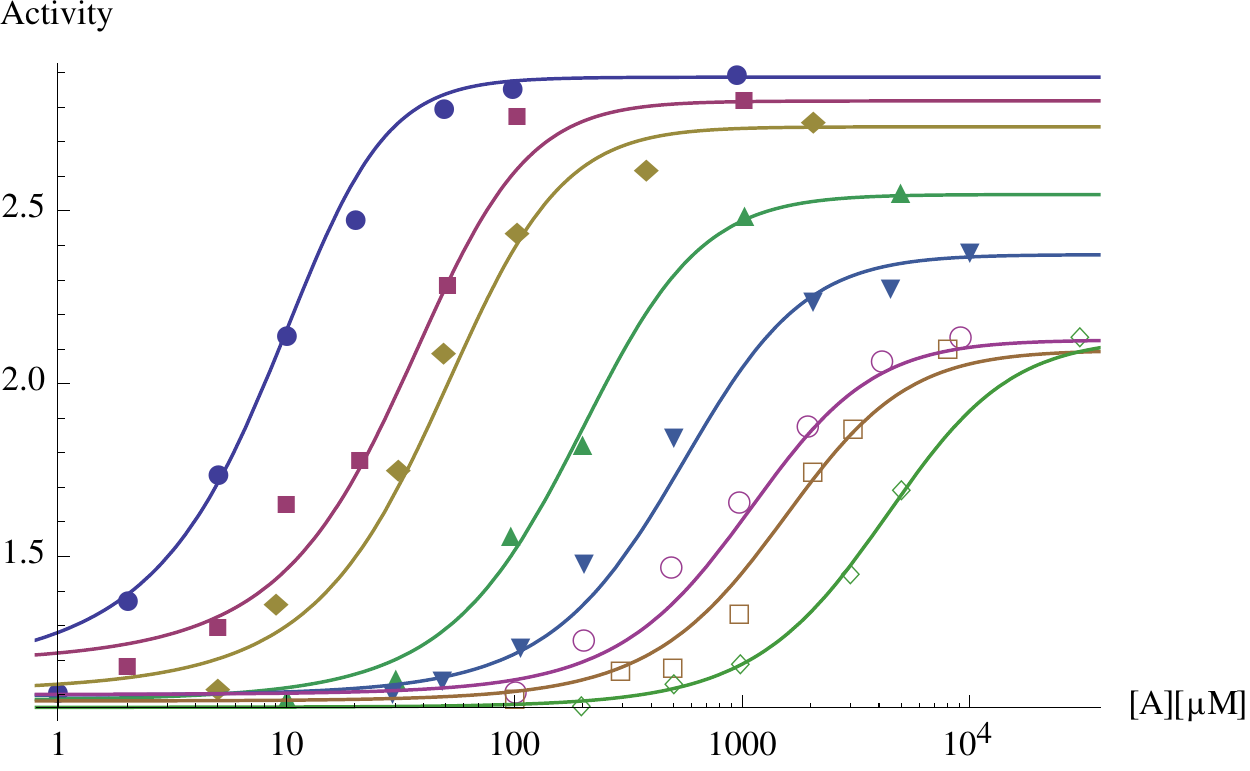}
\caption{Upper panel: stochastic YES gate, achieved through the statistical mechanical formulation of the allosteric monoreceptor-monoligand complex under assumptions $\mathcal A$ and tested on \emph{E. colii} chemotaxis network response measured by fluorescence resonance energy transfer (FRET) to decreasing concentrations (in mM) of $\alpha$-methylaspartate (MeAsp, [A]); data from \cite{Mello-PNAS2003}.
Lower panel: stochastic NOT gate, achieved under assumptions $\mathcal A'$ and tested on \emph{E. colii} FRET-measured chemotaxis network response to increasing concentrations (in mM) of MeAsp ([A]); data from \cite{Keymer}. See \cite{Mello-PNAS2003}, \cite{Keymer} for more details.
}
\begin{center}\label{YES-DataFit}
\end{center}
\end{figure}

\section{Conclusions: Merging statistical mechanics, logic and biochemistry}

\begin{figure}[htb!]
\begin{center}
\includegraphics[scale=0.2]{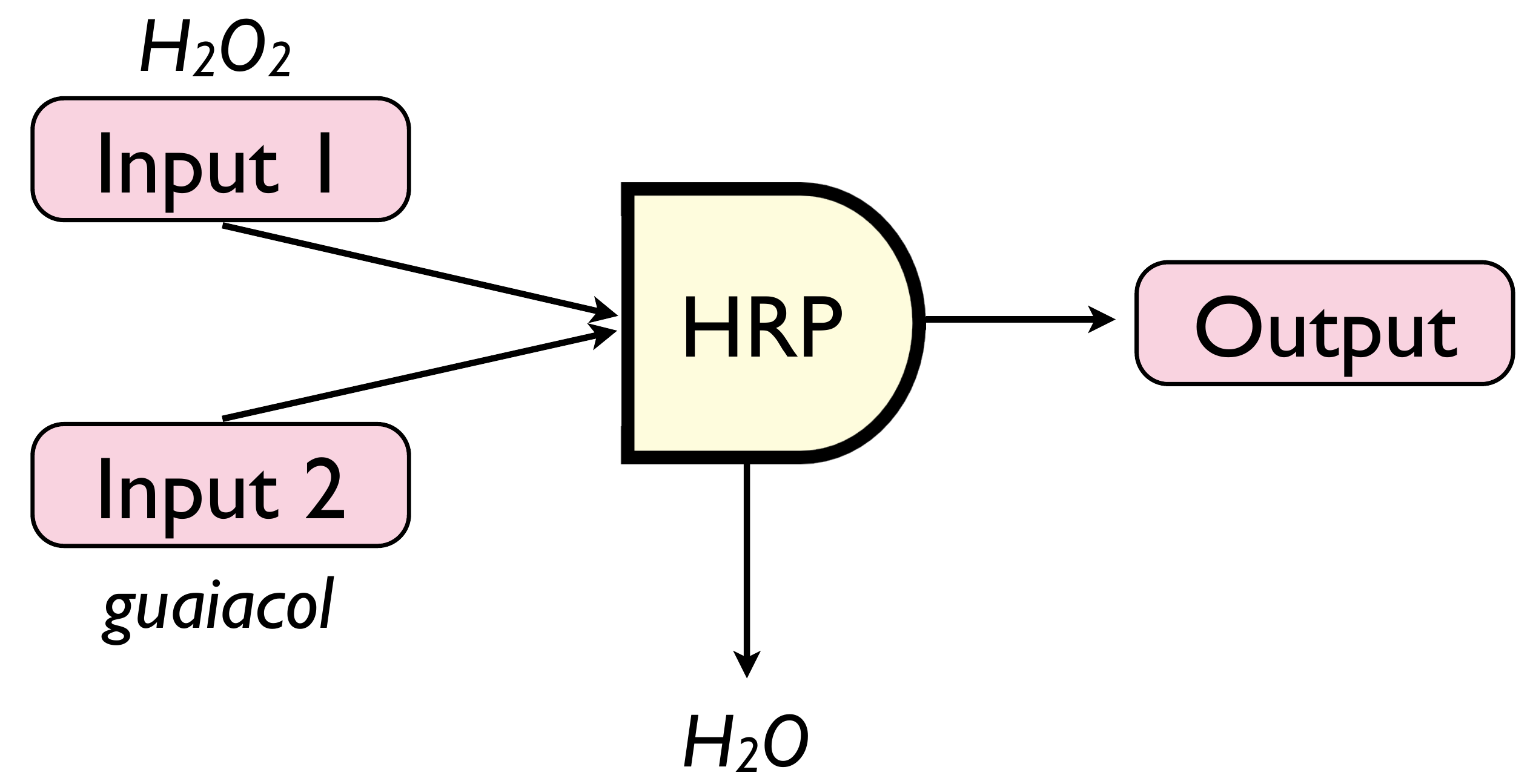} \ \ \ \ \includegraphics[scale=0.2]{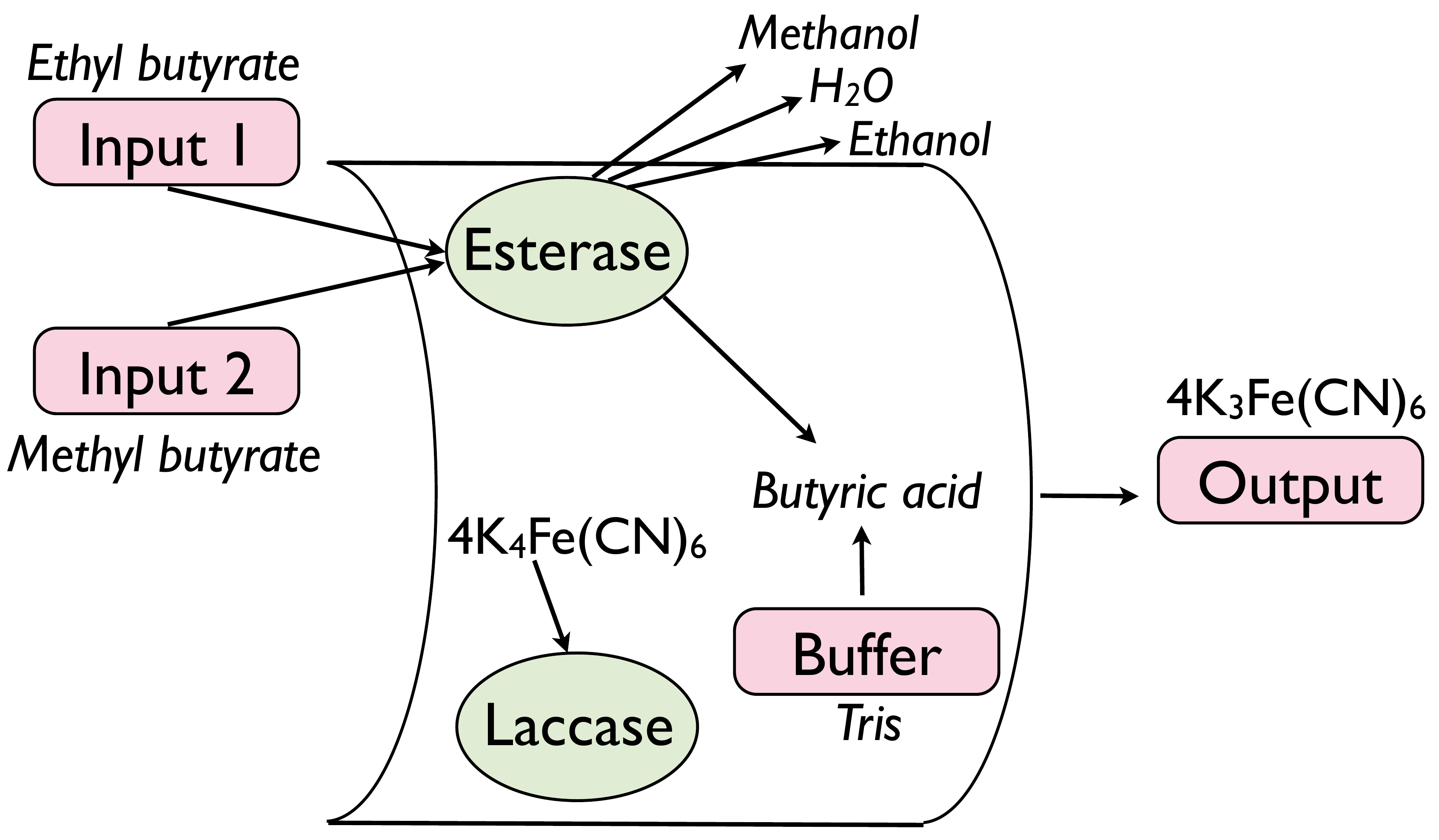}
\caption{Schematic representation of the gates from a biochemical perspective. Upper panel: The stochastic AND gate is shown as a biocatalytic process. The two inputs are $\mathrm{H}_2 \mathrm{O}_2$ and one out of three chromogens (ABTS, ferrocyanide, guaiacol) -only the latter is illustrated-. Signal processing is biocatalyzed by HRP and the output measure optically as the amount of the oxidized chromogen. See \cite{katz-and} for more details. Lower panel: The stochastic OR gate is shown. It involves two enzymatic processes and a buffering part. The first enzyme is esterase, that reacts with ethyl butyrate or methyl butyrate (or both) biocatalyzing production of ethanol and methanol, respectively. Butyric acid is a byproduct of the process and, as its production lowers the $pH$ of the system, further a buffer is added. The product of the process is measured by absorbance  at $\lambda  =
420$ nm  using a UV-2401PC/2501PC UV-visible spectrophotometer with a TCC-240A temperature controller holder. See \cite{katz-or} for more details.
}
\label{fig:Katz}
\end{center}
\end{figure}

\begin{figure*}[htb!]
\includegraphics[scale=0.35]{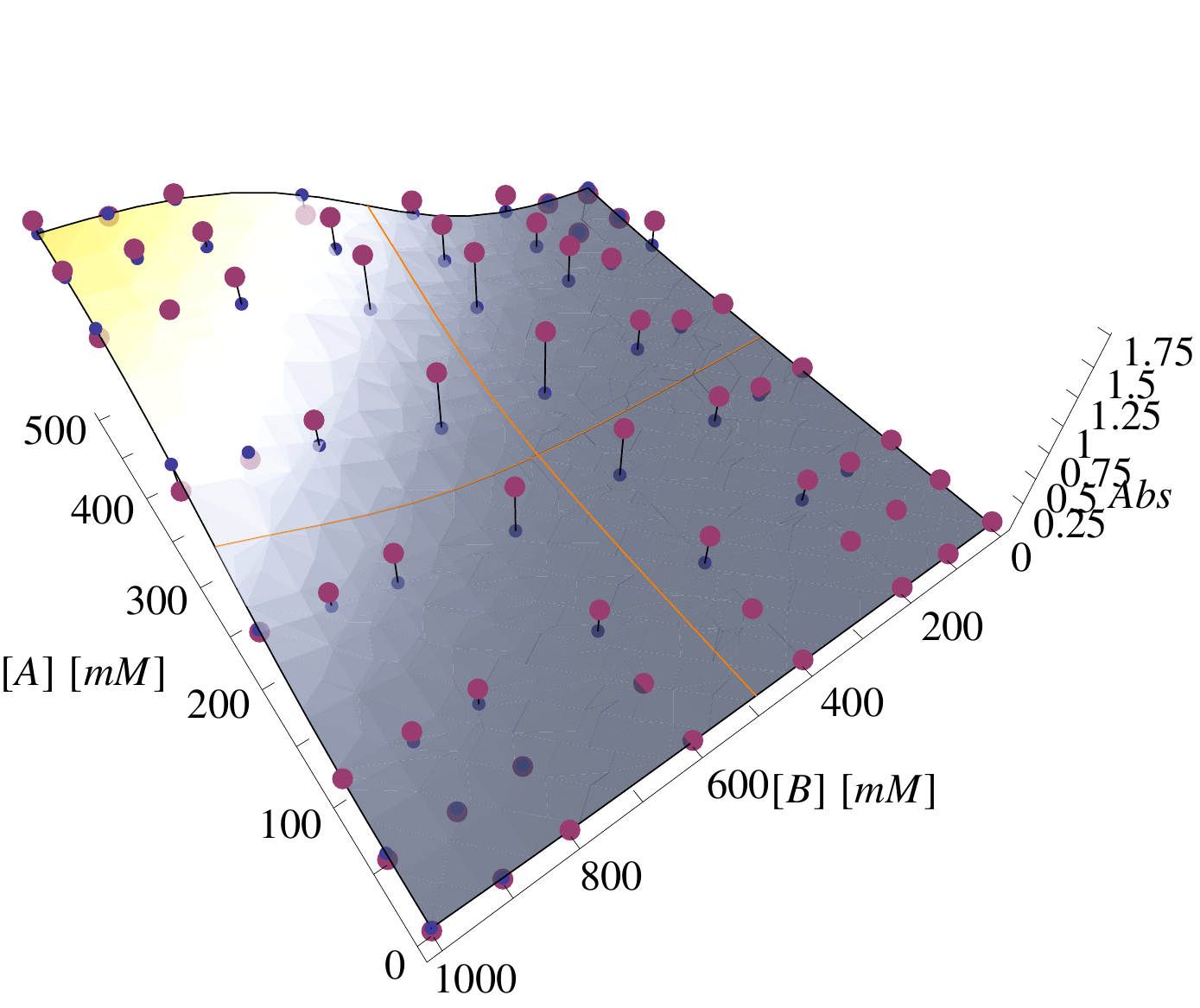}
\includegraphics[scale=0.35]{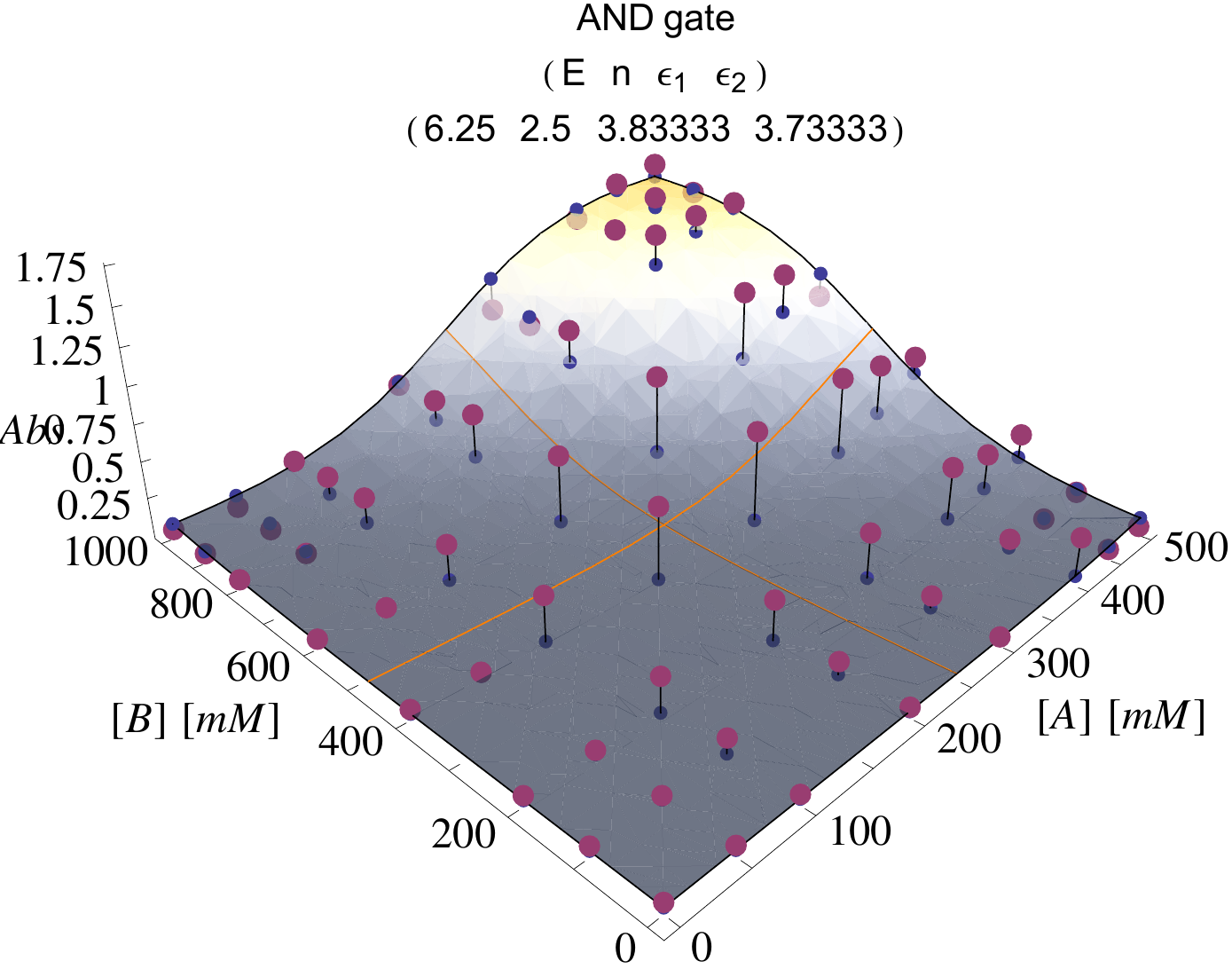}
\includegraphics[scale=0.35]{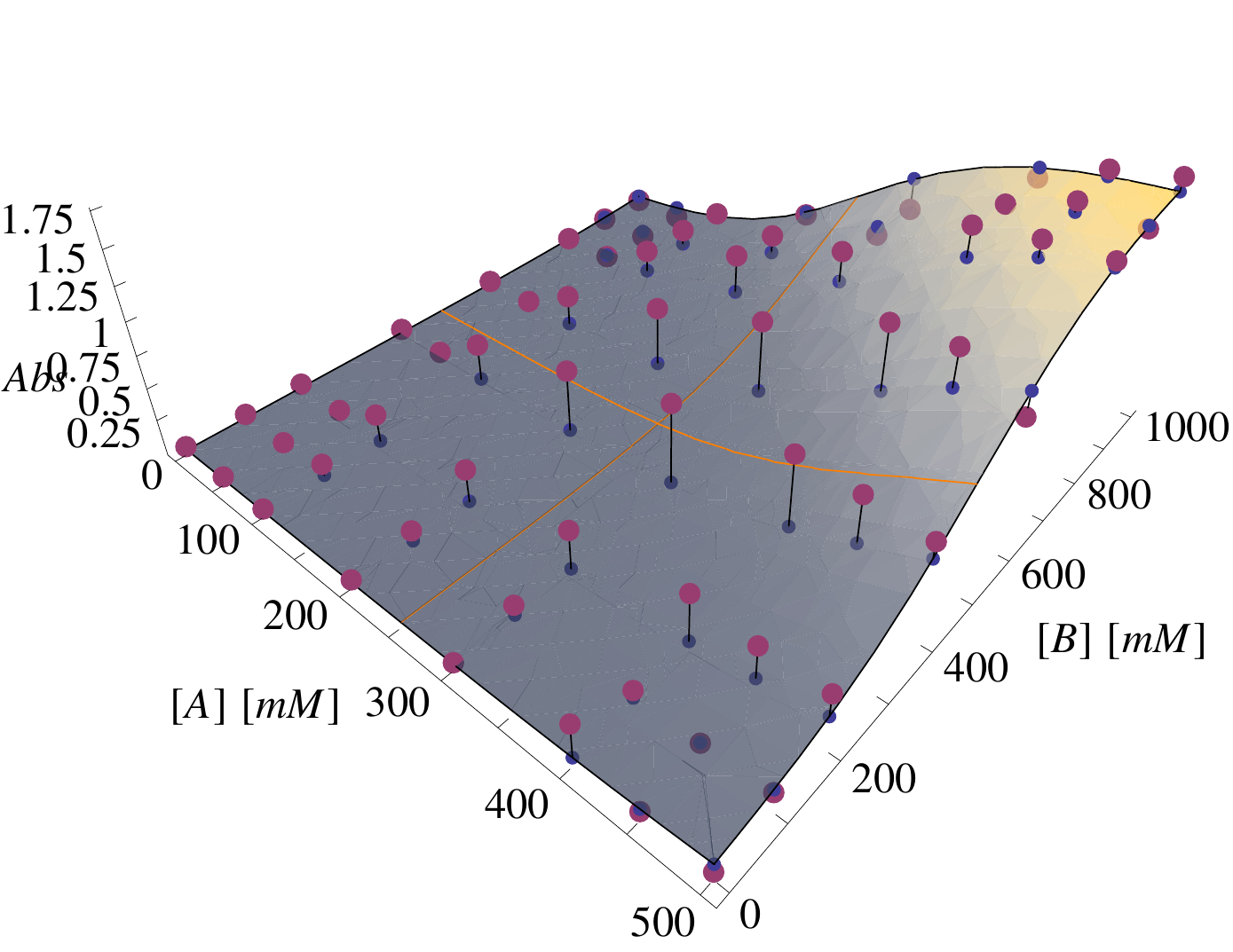}
\caption{The stochastic AND gate has been realized by two inputs constituted by $\textrm{H}_2\textrm{O}_2$ ([A]) and guaiacol ([B]) as the chromogen, while \textsc{l}-ascorbic acid [Asc](0) = 120 $\mu$M was used for filtering; the signal processing
was biocatalyzed by HRP, as sketched in Figure 7 (upper panel). The output was measured optically as the amount of the oxidized chromogen. Bullets represent experimental data \cite{katz-and}, whereas the surface
represents the best fitting according to eq.~\eqref{Pand}.}
\label{fig:AND-Fittato}
\end{figure*}

\begin{figure*}[htb!]
\includegraphics[scale=0.35]{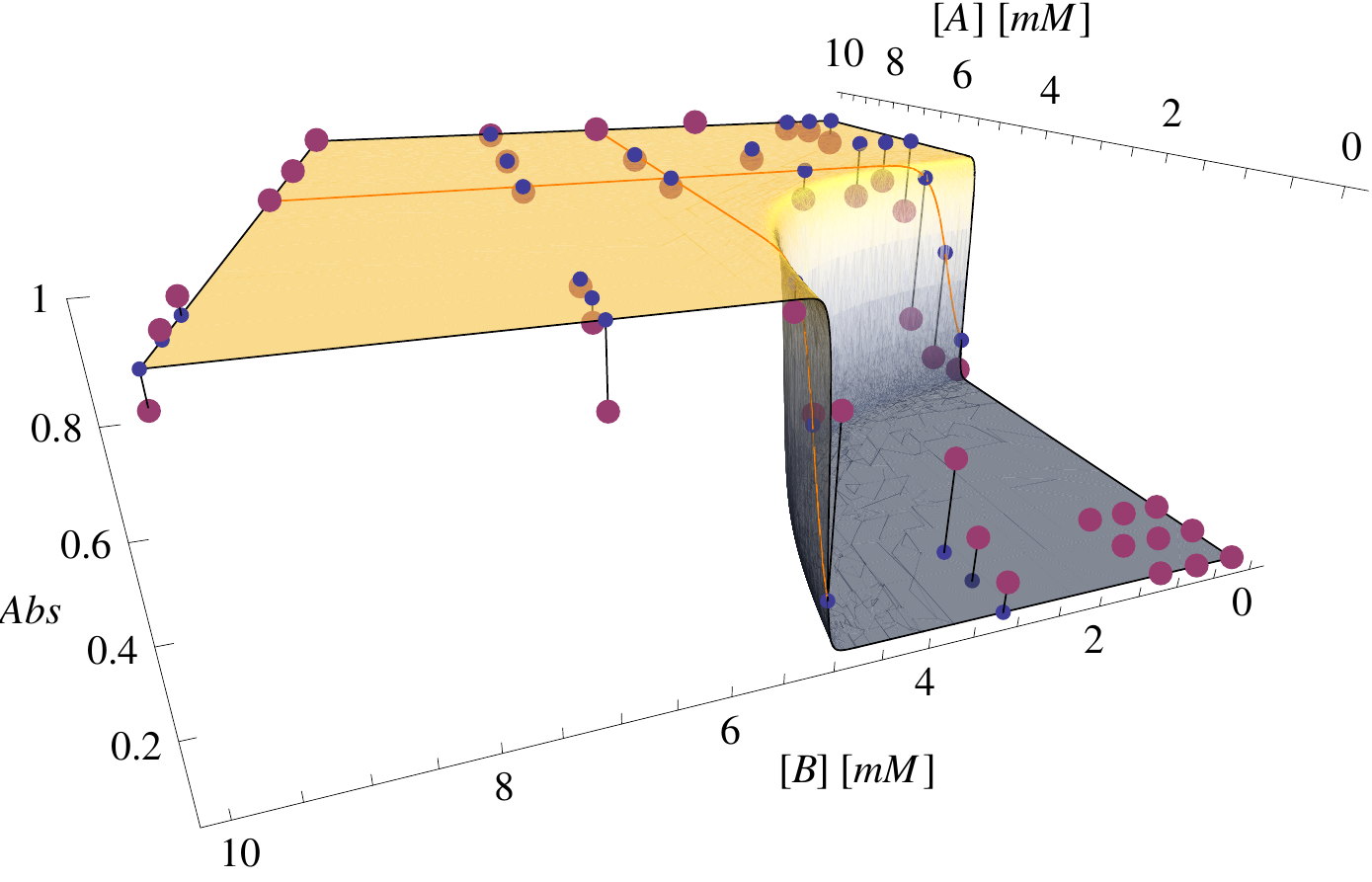}
\includegraphics[scale=0.35]{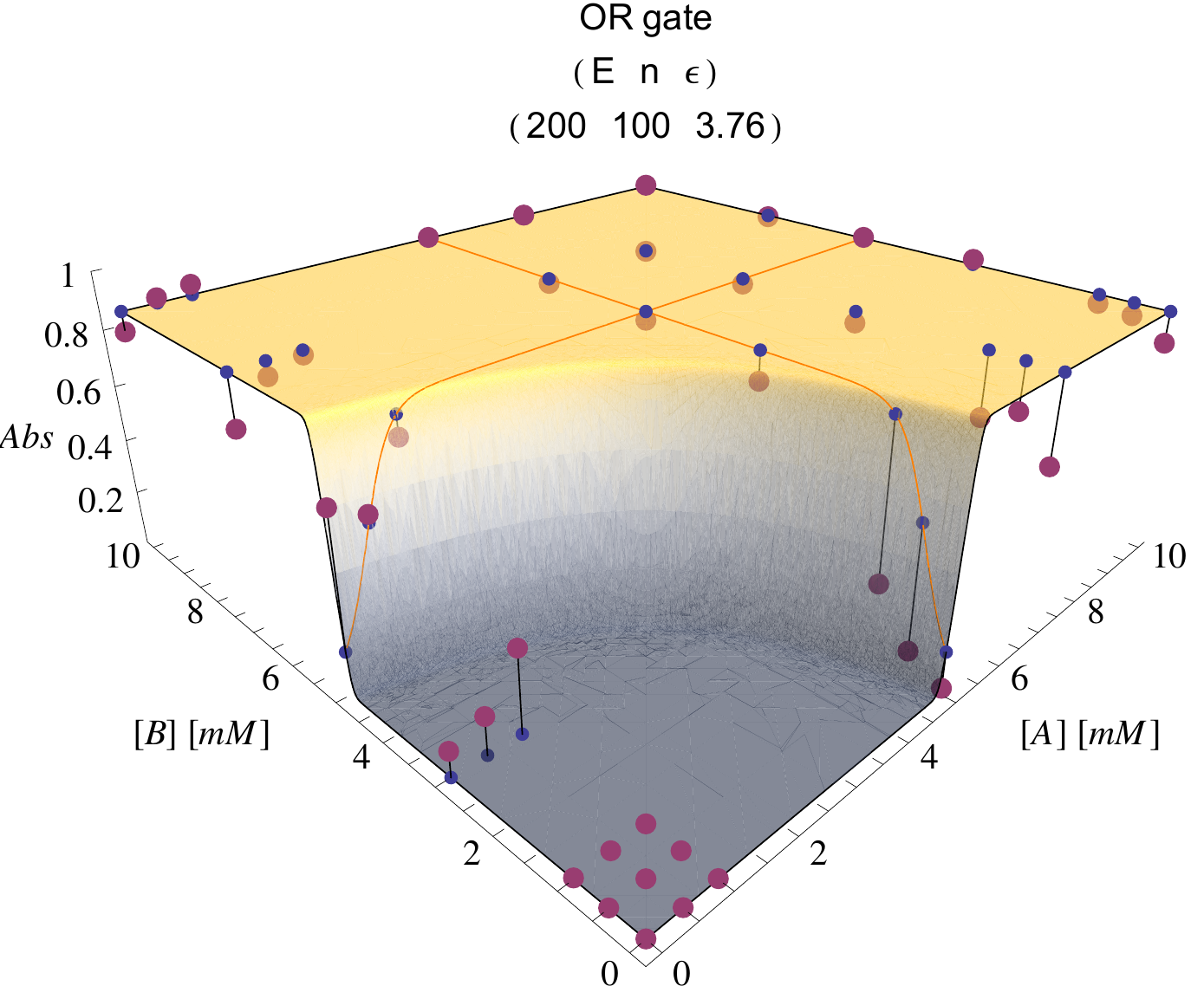}
\includegraphics[scale=0.35]{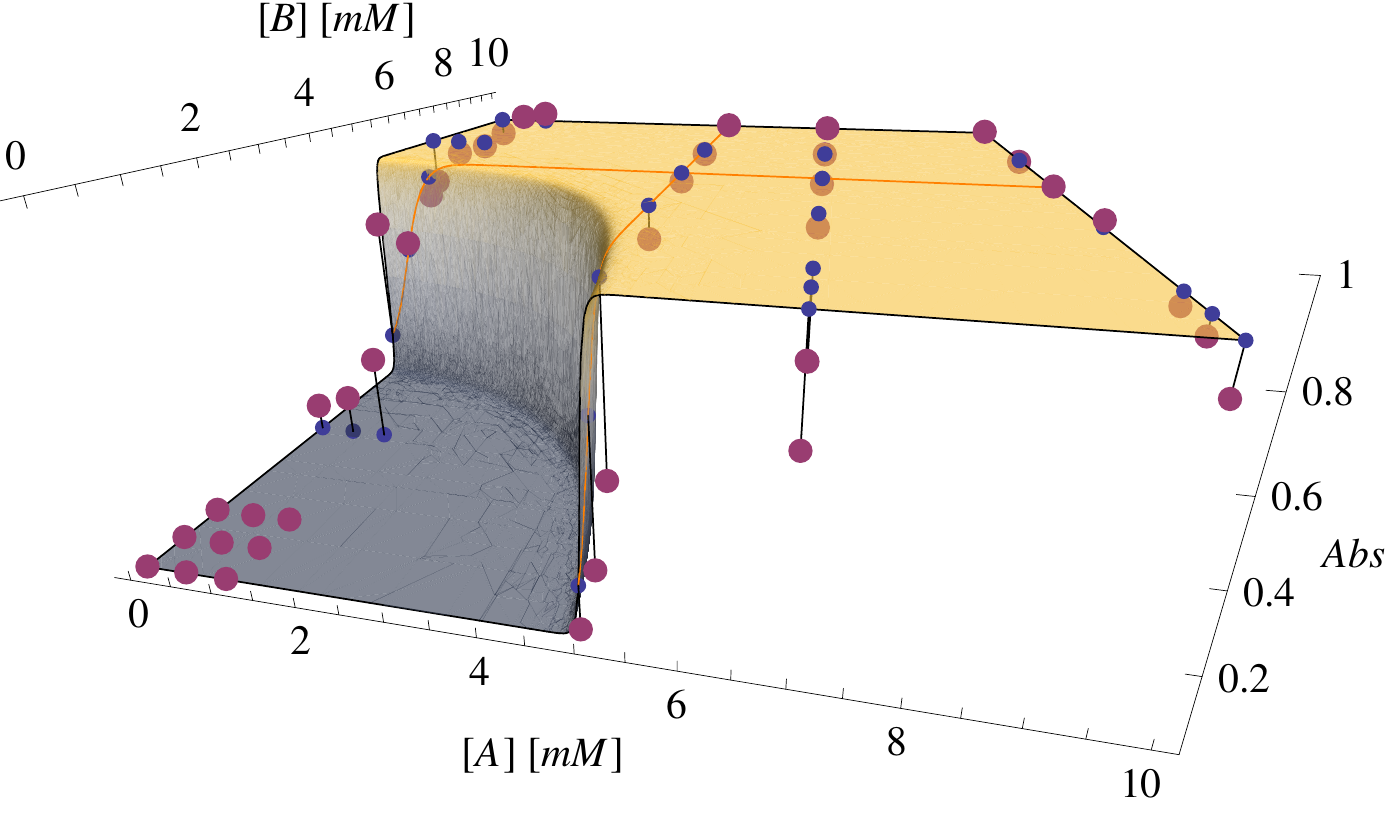}
\caption{The stochastic OR gate has been realized in two coupled steps involving enzymatic processes as sketched
in Figure~\ref{fig:Katz}: first enzyme is esterase, that reacts with ethyl butyrate ([A]) or methyl butyrate ([B]), or both, catalyzing production of ethanol and methanol, respectively. Butyric acid is a byproduct of the process.
To set the gate, the physical zeros of the signals have been fixed experimentally to convenient input values (ethyl butyrate $10$~mM and  methyl butyrate $10$~mM).
Bullets represent experimental data \cite{katz-or}, whereas the surface represents the best fitting according to eq.~\eqref{pAor}.
}
\label{fig:OR-Fittato}
\end{figure*}

We can finally test the predictions of the theory over the \emph{in vitro} experiments carried on both single-input and two-input (see Fig.~\ref{fig:Katz}) (bio)-logic gates and obtain our conclusions. Since the variable $h$ and parameters $n, E, \eps_1,\eps_2$ are dimensionless, any linear rescaling of the function $p_A$ is allowed  that suitably fits the data and whose choice is discussed below.

\subsection{Unary operators}
In the YES case (data from \cite{Mello-PNAS2003}), the opportune $y$-rescaling is obtained for each data set $D_k$ by considering the function $r_A^k\doteq (\max D_k-\min D_k)p_A + \min D_k$. In order to compensate the logarithmical progression of the axis, the $x$-rescaling (which is effectively linear, but conveyed on a log scale) is of the form $r_A^k(h)\doteq (\max D_k-\min D_k)p_A\left(h^m\right) + \min D_k$ where $m= m_k$ is opportunely depending on $k$.
The displayed function is $r^{\textrm{eff},k}_A$, which is the same as $r_A^k$, but varying parameters $n$, $E^\textrm{eff}\doteq 2n+k$, $\eps^\textrm{eff}\doteq 2E/n+\ell$. Consistently with scaling equations \eqref{eq:qui}, \eqref{eq:Escal}, $k$ varies within $\pm3.4\%E$ and $\ell$ within $\pm11.25\%\eps$.

In the NOT case (data from \cite{Keymer}) the opportune $y$-rescaling is obtained by plotting precisely the function $r_A^k\doteq (\max D_k-\min D_k)p_A + \min D_k$ with the same $x$-rescaling as in the YES case.
In order to show how precise the fitting is (\emph{after} suitable log-lin rescaling), the best fit is obtained by considering $r_A^k$ as a function of $n=n_k$ only, while $E_k=2n_k$ and $\eps=2E_k/n_k$, according to the assumptions, thus the fit is practically achieved with one degree of freedom only.

We emphasize that, in both cases, the fit may be improved by data extrapolation of maximal (minimal) values for the range of $r_A^k$ which are strictly higher (lower) than the maxima (minima) of $D_k$. 

\subsection{Binary operators}

Given the $x_1x_2$-data grid $\{0, ..., M_{1}\}\times\{0, ..., M_{2}\}$, a (vertical) $y$-rescaling is required in order to match $1$ with the experimental maximum value of the activation parameter. In order to determine such value, a \emph{stable data set} $S$ is opportunely defined;
letting $\langle S_z\rangle$ be the mean $z$-value of the stable data set, we take it as a reliable value for the maximal experimental activation.
The opportune $y$-rescaling is therefore obtained by considering the function $r_A\doteq \tfrac{\langle S_z\rangle}{p_A(0,0)}p_A$, while the $x_1x_2$-rescaling is achieved by plotting $r_A(h_1,h_2)=\tfrac{\langle S_z\rangle}{p_A(0,0)}p_A\left(\tfrac{M_1}{\epsilon_1}h_1,\tfrac{M_2}{\epsilon_2}h_2\right)$


In the OR case, the \emph{stable data set} is taken to be the data set in the $[8,10]$ mM $\times [8,10]$ mM region. The best fit is obtained by varying parameters $n$, $k$ and $\ell$, where the plotted function is an \emph{effective} $r_A$ function defined as $r_A^\textrm{eff}$, a function of $n$, $E^\textrm{eff}\doteq 2n+k$, $\epsilon^\textrm{eff}\doteq\tfrac{E}{n}+\ell$. Consistently with scaling equations \eqref{eq:qui}, \eqref{eq:Escal}, $k$ varies within $\pm 1 \% E$ and $\ell$ within $\pm 6.25\% \epsilon$.

In the AND case, the \emph{stable data set} is taken to be the data set in the $[400,500]$ mM $\times [800,1000]$ mM region. The best fit is obtained by varying parameters $n$, $k$, $\ell_1$ and $\ell_2$, where the plotted function is an \emph{effective} $r_A$ function defined as $r_A^\textrm{eff}$, a function of $n$, $E^\textrm{eff}\doteq 2n^2+k$, $\epsilon_{1,2}^\textrm{eff}\doteq\tfrac{2}{3}\tfrac{E}{n}+\ell_{1,2}$. Consistently with scaling equations \eqref{eq:Escal2}, \eqref{eq:qui2}, $k$ varies within $\pm 3 \% E$ and $\ell_{1,2}$ within $\pm 15\% \epsilon$.

Results are shown in Fig.~\ref{fig:AND-Fittato}, for the stochastic AND, and Fig.~\ref{fig:OR-Fittato} for the stochastic OR.

%
%

\section{Methods}\label{sec:methods}
In this section we discuss two major aspects of our work: the scaling assumptions and the role of \emph{allosteric cooperativity} within the model.

\subsection{Scaling assumptions}\label{ssec:Scal}
As assumption sets $\mathcal A, \mathcal A'$ only affect the sign of parameters $E$, $\eps$ and of the variable $h$, we cannot expect every choice of these quantities to yield a realistic behaviour from a biophysical viewpoint. Particularly an effective range of the variable $h$ as well as some reasonable scaling properties for $E$ and $\eps$ are to be determined, most likely depending on $n$.

The first issue can be solved independently of the case considered ($MM$, $MD$, $DD$).
As evidenced in Tab.~\ref{tab1}, for assumptions $\mathcal A$ it is $e^{-h}=[S]/K_I$ and, being $h$ positive, activation enhancement $[S]/K_I$ is dimensionless and ranging in $[0,1]$, thus, it may be considered as a \emph{percent molar concentration} of the ligand $S$.
Also, we expect that there exists a numerical (percent) value for the ligand concentration, below which the receptor activity is unaffected (see e.g. \cite{Linari-Biophys2004}). We refer to this threshold value as $\tau$ and, according to Tab.\ref{tab1}, this also determines the significance range of $h$ as
\beno
0<h<-\log \tau,
\eeno
which reliably limits the range of the dissociation energy to finite values. As $\tau$ determines the receptor sensitivity with respect to its activity, it is reasonably expected that $\tau\approx K_A/K_I$; in fact such inverse proportional dependence of $\tau$ with respect to $K_I$ is consistent with increasing monotonicity of $h$ with respect to $K_I$ (consistently with assumptions (\ref{assc}), (\ref{assf})).

Moreover, from Tab.~\ref{tab1}, $\tau\approx e^{-\eps}$, whence a reliable significance range for $h$ is
\be \label{hbioscaling}
0<h<\eps.
\ee

Dually, for assumptions $\mathcal{A}'$ it is $e^h=K_I/[S]$ and the same conclusion follows that $K_I/[S]$ may be considered as a \emph{percent molar concentration} of the ligand $S$. As for $\tau'$ we have $\tau'\approx K_I/K_A$ (following from assumptions $(c')$, $(e')$), yielding
\beno
-\eps<h<0.
\eeno

Now we focus on the scaling of $E$ and $\eps$: in the following we address this matter separately for the case of one or two receptors, which have different nature. 

\subsubsection{Mono-receptor case: YES/NOT and OR/NOR gates}
We refer only to assumptions $\mathcal A$, since dual gates clearly scale in the same way.
Let us start considering the Mono-Mono case: given Eq.~\ref{pAyes}, we can define $\overline{h}$ as the value of the dissociation energy such that $(p_A)_{MM}(\overline{h})=1/2$, which implies
\beno
e^{-E}\left(1+e^{\eps- \overline{h}}\right)^n = \left(1+e^{-\overline{h}}\right)^n.
\eeno

On the other hand, the active ($a=1$) and saturated ($\sigma=(1)$) state is an extremal state of system corresponding to minimum entropy. As a result, it is mathematically reasonable that
\beno 
H(a=1, \sigma=1, \overline{h})  = E - n\eps + \overline{h}n =0.
\eeno

From the previous two equations we have
\be\label{eq:qui}
\overline{h}=\tfrac{E}{n},\ \ \eps=\tfrac{2E}{n},
\ee

The same conclusion can be drawn independently following another route:  according to the constraint  (\ref{hbioscaling}), the maximum value attainable by the Hamiltonian (\ref{yeshamiltonian}) is $E$ and it corresponds  to an active state with $h = \eps$; on the other hand, the minimum value attainable is $E - n \eps$, corresponding to $h=0$ and a fully occupied state.
Imposing the range interval for the energy $[E - n \epsilon, \ E]$ to be symmetric around $0$ it must then be $E - n \epsilon = -E$, namely $\eps =2E/n$.
Finally we observe that $E$ depends only on the receptor, therefore in the presence of a single receptor-type it must be $E \propto n$ in view of the linear
extensively of thermodynamics; direct verification shows that
\be\label{eq:Escal}
E \approx 2n
\ee

best fits our purpose.

Scaling assumptions for the OR gate are derived imposing that the behavior of the function $(p_A)_{MD}$ recovers that of $(p_A)_{MM}$ when one of the ligands
is absent (that is, when $h_2 \to \infty$). If we carry out the calculation, we find that
\beno
(p_A)_{MD} \Big \rvert_{h_2 = \infty} = (p_A)_{MM}
\eeno
so the scaling for $E$ and $\eps$ must be the same of the previous one in order to be consistent.

\subsubsection{Double-receptor case: AND/NAND gates}
This case is different from the Mono-receptor one mostly because of the non-linear scaling of $E$: since the receptors are dimeric, their response
must be linear with respect to each functional monomer; consequently $E \propto n^2$, and again we see directly
that the proper scaling is achieved by
\be\label{eq:Escal2}
E \approx n^2.
\ee

As far as the scaling of $\eps$ is concerned, we proceed in the same way as we did for the OR gate, and argue that posing $h_2 = 0$ (strong presence of
one ligand) must logically recover the behavior of $(p_A)_{MM}$ from $(p_A)_{DD}$. In this case, however, we do not find an exact identity, but
we can rearrange the result to look like what we expect. In fact, we have
\bano
(p_A)_{DD}\bigg|_{h_2=0} = \frac{ \frac{e^{-2E} (1 + e^{\epsilon})^n }{2^n} (1+ e^{\epsilon - h_1})^n  }{(1 + e^{-h_1})^n + \frac{e^{-2E} (1 + e^{\epsilon})^n }{2^n} (1 + e^{\epsilon -h_1})^n },
\eano
so, setting $e^{-E^{\mathrm{eff}}} = \tfrac{e^{-2E} (1 + e^{\epsilon})^n }{2^n}$ and $\eps = 2E^{\mathrm{eff}}/n$ we obtain
\be\label{eq:qui2}
\eps = 4E/3n.
\ee

\subsection{The role of allosteric cooperativity}
Now we want to make clear where the differences between the classical cooperativity and the MWC-like one,
known in the Literature as {\em allosteric cooperativity} (see e.g., \cite{hev, KNF}), reside.
This difference can be investigated directly from a mathematical and logical point of view by comparing
the plots of the AND gate and of the OR gate.

\subsubsection{OR gate: classical cooperativity}

We here discuss why and how the OR gate, that can be handled by a one-body statistical
mechanical Hamiltonian (eq.~\eqref{orhamiltonian}), does manifest a (roughly standard) cooperative behavior. The OR Hamiltonian is indeed a rigged one-body expression: cooperativity (meant as produced by a term  $\sim J \sigma \sigma$, see eq.~\eqref{coopera1}) is nested within the definition of the OR Hamiltonian coded in eq.~\eqref{orhamiltonian}, hidden inside the request  $I\cap J= \varnothing$. It is in fact possible to infer from this constraint that, in order to obtain the correct ensemble $K$ of the indices of the occupied binding sites, it is alternatively possible to introduce two subsets $I'$ and $J'$ where only the condition $I',J'\subset N$ is left to be respected: the price to pay for this simplification, however, is in writing the ensemble $K$ as $K=I'\sqcup J'\setminus I'\cap J'$, instead of $K=I\cup J$.
Such way of writing the OR constraints (which is nothing but a reformulation of the Inclusion-Exclusion Principle) makes explicit the presence of the cooperative term which turns out to be exactly $\sum_{k\in I'\cap J'}\sigma_k$. The latter can be rewritten as $\propto \sum_{i,j}J_{ij}\sigma_i\sigma_j$ (for some positive coupling $J$) because $\sigma_i\sigma_j=1$ if and only if both $\sigma_i=1$ and $\sigma_j=1$.
As a further check of the latter statement it is to be noticed that the presence of a quadratic growth term accounting for proper cooperativity may be deduced by the circular edge of the upper plateau (Fig.~4).

\subsubsection{AND gate: allosteric cooperativity}

In a real cooperative system there is a mutual enhancement of the activation probability; conversely,
the AND gate lacks such a mutual enhancement, and the presence itself of both the ligands is simply necessary for activation, or, in other words,
it is possible to (biochemically) realize an AND gate only when a (significant, that is {\em at high concentration}) amount of both ligands
is present, independently of the percent concentration relative to any of them.
Since the AND Hamiltonian (eq. (\ref{andhamiltonian})) results only from the juxtaposition of two YES Hamiltonians, it is truly one-body:
this fact is fully consistent with the linear edge of the upper plateau in the AND plot (Fig.5).
%

Note that, if instead of an allosteric mechanics (hence with the activation parameter $a$ and with two different conformational states $R, T$),
we adopted a classical (i.e. not-allosteric) cooperative Hamiltonian for the system, we would write
\ba\label{coopera1}
H(\sigma,\tau) &=&   H_{12}(\sigma,\tau) + H_1(\sigma) + H_2(\tau) \\
\nonumber
&=& - J \sum_{i,j}^{n_1,n_2}\sigma_i \tau_j + h_1 \sum_{i=1}^{n_1}\sigma_i + h_2 \sum_{i=1}^{n_2}\tau_i,
\ea
where $J$ is a scalar parameter tuning the reciprocal enhancement.

Comparing eq.~\eqref{andhamiltonian} and eq.~\eqref{coopera1} we see that they would be equivalent if we could write
$\epsilon_1 \equiv \epsilon_1(\tau)$ and $\epsilon_2 \equiv \epsilon_2(\sigma)$ but, as $\log \epsilon = K_I/K_A$,
then $\epsilon_1$ and $\epsilon_2$ are constant dependent on the species making up the system but independent of their bounding state,
that is, $\epsilon_1 \neq \epsilon_1(\tau)$ as well as $\epsilon_2 \neq \epsilon_2(\sigma)$ (see \cite{ABBDU-ScRep2013} for classical cooperativity).
Therefore, we cannot express the Hamiltonian \eqref{andhamiltonian} as a two-body system, and this codes for the \emph{allosteric} nature of this gate.

We perform now a brief mathematical analysis of the above mentioned shape of the AND plot (from here on referred to as a ``\emph{cut}''): a simple calculation shows that
$\partial_{h_1}(p_A)_{DD}\Big\lvert_{h_1, -h_1+\eps/2}=\partial_{h_2}(p_A)_{DD}\Big\lvert_{h_1, -h_1+\eps/2}$, which states that
the \emph{cut} is in fact corresponding to the straight line $h_2=-h_1+\eps/2$ (the symmetric angular coefficient simply remembers the choice $\eps_1=\eps_2$).
Furthermore, it is possible to prove that the slope $m$ of the line
projection on the $h_1,h_2$-plane is in fact $m \approx \eps_1/\eps_2$. It follows that the case $\eps_1=\eps_2=\eps$ is the one best fitting the
expected plot of the logical operator. On the contrary, by taking limits for either $\eps_1\rar\infty$ or $\eps_2\rar\infty$, one recovers
the YES$_2$ (respectively YES$_1$) as a projection on the (orthogonal) axis.

As a consequence of this discussion,
there is no contradiction between the observed behavior of the AND gate and a statistical mechanical scaffold built on a one-body
Hamiltonian because, effectively, the AND gate does not display a classical cooperative behavior, but, rather, it has its reward by a useful
{\em alliance} among ligands, alliance that we call {\em allosteric cooperativity}.

\section*{Acknowledgments}

This work is supported by Gruppo Nazionale per la Fisica Matematica (INdAM) through Progetto Giovani (Agliari, 2014).

\section*{Author contributions}

EA and AB proposed the theoretical research and gave the lines to follow. LDS and MA made all the calculations [solving all the related problems (e.g., suitable scalings of the parameters, etc.)] and fitted the theory over the data. The latter come from the experimental route that has been completely guided by EK.
\newline
All the authors wrote the paper in a continuous synergy.

\section*{Additional information}

The authors declare no competing financial interests.


\bibliographystyle{plain}
\bibliography{biblio_CK}

\end{document}